\documentclass[onecolumn]{DDLabJournalPaper}

\labname{Dibakar Datta Laboratory}
\labshortname{DDLab}
\institutionshort{NJIT}

\labwebsite[https://dibakardatta.net/]{www.dibakardatta.net}

\labcodeurl{https://dibakardatta.net/code/}

\citationauthors{D. Datta \emph{et al.}}

\citationauthors{D. Datta}          

\makeatletter
\patchcmd{\DDJ@titleblock}{\LARGE\bfseries}{\Large\bfseries}{}{}
\patchcmd{\DDJ@titleblock}{{\normalsize\DDJ@renderauthors\par}}%
                          {{\large\DDJ@renderauthors\par}}{}{}
\makeatother

\newcommand{\ion}[2]{\mbox{#1$^{#2}$}}
\newcommand{\TierGoal}[1]{\par\noindent{\itshape\color{DDNavy}#1}\par\smallskip}
\newcommand{\Lesson}[1]{\textbf{#1}\par\smallskip\noindent}

\usepackage{newfloat}
\DeclareFloatingEnvironment[name=Scheme,listname={List of Schemes},fileext=los]{scheme}
\crefname{scheme}{Scheme}{Schemes}
\Crefname{scheme}{Scheme}{Schemes}

\newcommand{\figw}{0.92\linewidth}

\manuscripttype{Perspective}
\venue{Preprint}
\preprintid{arXiv:2608.02858}
\manuscriptversion{v2}

\title{Physics-Informed and Knowledge-Driven Generative AI for Autonomous
       Discovery of Porous Oxide Energy Materials: Opportunities and
       Challenges}
\shorttitle{Physics-informed generative AI for porous oxide energy materials}

\DDAuthor{\textbf{Dibakar Datta}}{}
\citationyear{2026}

\DDAffiliation{}{Dibakar Datta Laboratory (DDLab), Department of Mechanical
  and Industrial Engineering\\
  New Jersey Institute of Technology (NJIT), Newark, NJ 07102, USA}

\correspondence{\href{mailto:ddlab@njit.edu}{ddlab@njit.edu};
  Phone: +1-973-596-3647}
\date{\today}

\paperabstract{%
The discovery of next-generation energy-storage materials is increasingly
limited not by the lack of computational methods, but by the complexity of the
underlying design problem. Porous transition-metal oxides represent a
particularly challenging class of battery materials because their performance
emerges from strongly coupled interactions among crystal chemistry, pore
architecture, ion transport, electrochemistry, electro-chemo-mechanics,
synthesis, manufacturing, and full-cell operation. Recent advances in
generative artificial intelligence (AI), including diffusion models, crystal
variational autoencoders, and large language models, have demonstrated
remarkable capabilities for generating chemically plausible crystal
structures. However, current approaches remain largely focused on
crystallographic validity and thermodynamic stability, while treating many of
the physical, electrochemical, and engineering requirements governing
practical battery performance as post-generation evaluation criteria. In this
perspective, we argue that the next generation of AI-enabled materials
discovery must move beyond crystal generation toward physics-informed,
application-aware, and synthesis-aware inverse design. Using porous oxide
electrodes as a representative materials platform, we first examine why these
materials constitute a distinctive inverse-design problem and discuss the
limitations of conventional experimental, computational, and AI-driven
discovery strategies. We then analyze the capabilities and limitations of
first-generation generative AI through a representative case study. Building
upon these observations, we propose a seven-tier physics-informed
inverse-design framework that progressively integrates chemical validity,
thermodynamic viability, transport functionality, electrochemical
performance, electro-chemo-mechanical durability, electrode and cell
compatibility, and manufacturability into a unified design hierarchy. We
further identify the \emph{Missing Data Problem} as a fundamental bottleneck
limiting application-aware generative AI and introduce an autonomous
knowledge-generation framework supported by a Porous Oxide Energy Materials
Ontology and a continuously evolving Porous Oxide Energy Materials Knowledge
Base. These concepts provide the semantic and scientific foundation for
Synthesis-Aware, Closed-Loop Autonomous Discovery, in which literature mining,
multiscale simulations, autonomous experimentation, and knowledge generation
operate within a continuously learning scientific ecosystem. Although this
perspective focuses on porous oxide materials for next-generation energy
storage, the conceptual frameworks proposed here extend naturally to many
other classes of functional materials. More broadly, we envision a transition
from AI systems that generate crystal structures to autonomous scientific
partners capable of integrating knowledge generation, inverse design,
computation, experimentation, and continual learning to accelerate materials
discovery.}

\begin{document}
\maketitle

\section{Introduction: From Porous Architectures to Autonomous Materials Discovery}
\label{sec:intro}

\subsection{Motivation}
\label{sec:motivation}

The global transition toward electrified transportation, renewable-energy
integration, portable electronics, aerospace applications, and grid-scale
energy storage is driving unprecedented demand for rechargeable batteries with
higher energy density, faster charging, longer cycle life, improved safety,
and lower cost \citep{chu2012opportunities}. Although lithium-ion batteries remain the dominant
electrochemical energy-storage technology, their continued development is
constrained by limitations in electrode materials, including sluggish
transport kinetics, structural degradation, limited operating-temperature
range, resource availability, and manufacturing scalability \citep{goodenough2013liion}.
Consequently, discovering next-generation electrode materials has become one
of the central challenges in modern materials science \citep{yu2021sustainable,periyasamy2026emerging}. Battery
performance (\cref{fig:mam}a) depends on far more than chemical composition
\citep{palacin2016batteries,meng2017advances}. Practical electrodes require simultaneous optimization of ion
transport, electronic conductivity, redox activity, structural stability,
mechanical durability, electrolyte compatibility, electrode architecture, and
manufacturability across multiple length scales, from atomic-scale ion
insertion and charge transfer to particle deformation, electrode transport,
and full-cell operation \citep{tian2021promises}. Designing materials that satisfy these
coupled requirements remains a formidable scientific and engineering challenge
\citep{shi2022challenges,yu2014designing,liu2019challenges}.

\subsection{Why porous oxides?}
\label{sec:why-porous}

\begin{figure}[tbp]
  \centering
  \includegraphics[width=\figw]{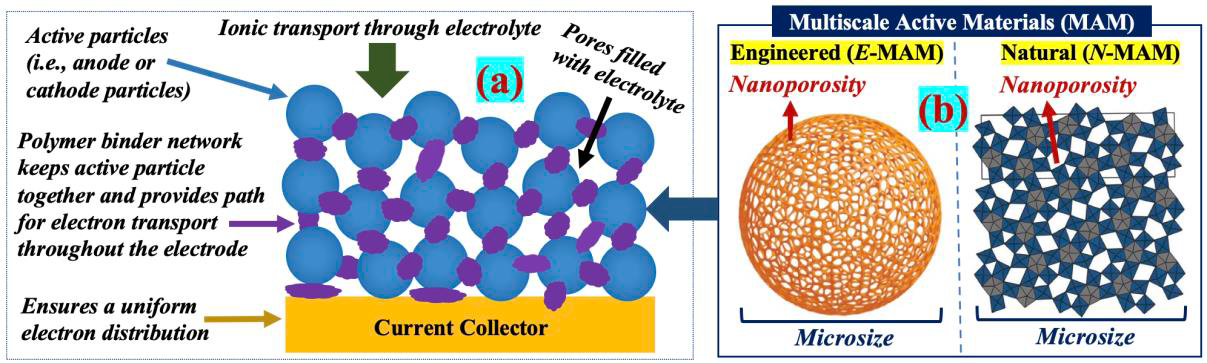}
  \caption{Multiscale active materials. (a) Basic design of a battery
  electrode. (b) Multiscale Active Materials (MAM) as electrode:
  ``engineered'' (E-MAM) and ``natural'' (N-MAM). Reprinted (adapted) with
  permission from ref.~\citep{datta2024ecm}. Copyright 2024 The Minerals, Metals \&
  Materials Society.}
  \label{fig:mam}
\end{figure}

\begin{figure}[tbp]
  \centering
  \includegraphics[width=\figw]{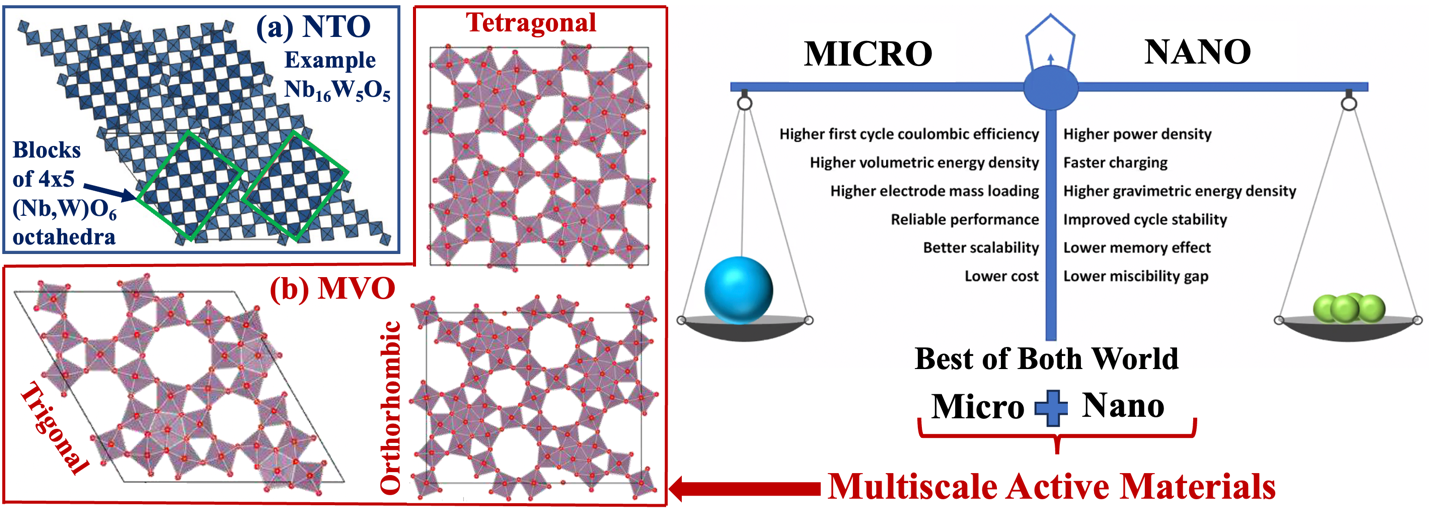}
  \caption{Examples of multiscale active materials (MAM). (a) Niobium
  tungsten oxides (NTO) and (b) molybdenum vanadium oxides (MVO) with various
  pore topologies. MAM combine the best of both worlds, micro and nano.
  Reprinted (adapted) with permission from ref.~\citep{datta2024ecm}. Copyright 2024
  The Minerals, Metals \& Materials Society.}
  \label{fig:mam-examples}
\end{figure}

One of the longstanding challenges in battery design is balancing the
advantages of microstructured and nanostructured electrodes \citep{jain2022nanostructuring,lucero2022roles}
(\cref{fig:mam,fig:mam-examples}). Micrometer-sized particles remain the
industrial standard because they provide high tap density, high
active-material loading, favorable volumetric energy density, and
compatibility with scalable manufacturing \citep{jain2022nanostructuring,attia2025challenges}. However, their long
diffusion distances often result in sluggish kinetics, concentration
gradients, stress accumulation, and mechanical degradation during cycling. In
contrast, nanoparticles shorten diffusion pathways and improve rate capability
but suffer from excessive surface reactions, electrolyte decomposition, low
tap density, particle aggregation, and increased manufacturing complexity
\citep{jain2022nanostructuring,datta2025electro,datta2014enhanced,datta2014defective,mukherjee2014defect}.

Porous open-framework oxides naturally bridge these length scales
(\cref{fig:mam}b, \cref{fig:mam-examples}) \citep{datta2024ecm}. Rather than replacing
microparticles with nanoparticles, they retain a microscale external
morphology while incorporating crystallographically defined nanoscale tunnels
and channels within the crystal framework \citep{jain2022nanostructuring,datta2024ecm}. This unique
combination of microscale processability and nanoscale transport makes porous
oxides natural multiscale active materials, combining the manufacturing
advantages of conventional microparticles with the transport benefits of
nanostructured electrodes (\cref{fig:mam,fig:mam-examples})
\citep{datta2024ecm,bhimani2023molybdenum}. Wadsley--Roth shear phases \citep{yang2021wadsley}
(\cref{fig:mam-examples}a), molybdenum--vanadium oxides
(\cref{fig:mam-examples}b), manganese oxides \citep{mahajani2026cobalt,vemuri2022oxygen}, and related
transition-metal oxides exemplify this architecture by providing
interconnected ion-transport pathways while maintaining structural integrity
during repeated electrochemical cycling \citep{datta2024ecm}.

These crystallographic transport networks make porous oxides particularly
attractive for emerging battery technologies, including multivalent-ion
\citep{lakhnot2022intercalation,anjan2026multivalent} (\cref{fig:adsorption}), fast-charging, and low-temperature
batteries \citep{kim2024enabling,kim2024innovative} (\cref{fig:lowT}). Their open frameworks facilitate
rapid ion transport, partially accommodate insertion-induced lattice
distortion, support multiple redox-active transition-metal centers, and
improve electrochemical performance under demanding operating conditions
\citep{jain2022nanostructuring}. At the same time, their remarkable structural diversity provides
an extensive design space in which channel geometry, pore topology, framework
density, and crystallographic connectivity can be tailored for different
battery chemistries \citep{datta2024ecm}. Porosity alone, however, is insufficient. The
practical performance depends on whether the pore architecture is chemically
and geometrically compatible with the intended charge carrier and operating
conditions \citep{datta2024ecm}.

\subsection{Why are porous oxides difficult to design?}
\label{sec:why-difficult}

Despite their promise, porous oxides are among the most challenging battery
materials to design because their performance emerges from multiple strongly
coupled physical phenomena rather than any single material property. Effective
porous oxides must simultaneously provide chemically stable crystal
structures, interconnected ion-accessible transport pathways, favorable
electronic conductivity, reversible redox reactions, mechanical robustness
during repeated cycling, and compatibility with the surrounding electrolyte
and electrode architecture. Optimizing one characteristic often compromises
another. Enlarging pore dimensions may improve ion accessibility but reduce
volumetric energy density, whereas increasing framework rigidity may enhance
structural stability while impeding ion transport and stress relaxation
(\cref{fig:mam-examples}) \citep{yang2021wadsley}.

These challenges become even more pronounced for multivalent-ion batteries
\citep{bhimani2023molybdenum,lakhnot2022intercalation,lakhnot2022reversible} (\cref{fig:adsorption}) based on \ion{Mg}{2+},
\ion{Ca}{2+}, \ion{Zn}{2+}, \ion{Al}{3+}, etc. Compared with \ion{Li}{+},
multivalent ions exhibit higher charge density and stronger electrostatic
interactions with both the electrolyte and host framework, leading to larger
desolvation barriers, slower solid-state diffusion, and greater
insertion-induced lattice distortion \citep{lakhnot2022intercalation}. Consequently, the success of
porous oxides depends on the coupled interactions among pore geometry,
transport pathways, electrochemistry, mechanics \citep{datta2024ecm,naik2023ecm}, and the
surrounding electrochemical environment. Porous oxide discovery therefore
represents a genuinely multidimensional inverse-design problem rather than
conventional crystal-structure optimization.

\subsection{Why conventional discovery is insufficient}
\label{sec:why-conventional-intro}

The discovery of porous oxide materials is challenged by the enormous
complexity of the accessible design space \citep{datta2024ecm}. Transition-metal oxides
exhibit diverse compositions, stoichiometries, oxidation states, polymorphs,
defect configurations, and crystallographic frameworks, each possessing
distinct transport pathways and electrochemical behavior. Exploring this
multidimensional landscape through sequential synthesis and characterization
is prohibitively slow, while exhaustive first-principles calculations
\citep{datta2024unlocking} become computationally intractable for complex oxide systems with
large unit cells, cation disorder, multiple insertion sites, and competing
polymorphs.

Machine learning and deep learning \citep{lecun2015deep} have accelerated materials
discovery by replacing selected high-cost calculations with predictive
surrogate models \citep{datta2024unlocking,sharma2022developing,chen2019graph}. However, these approaches remain
fundamentally evaluative: they predict the properties of structures that are
already known or manually generated rather than discovering entirely new
materials \citep{datta2023transferable,curtarolo2013highway}. Consequently, they do not address the more
fundamental challenge of identifying previously unknown crystal structures
that simultaneously satisfy multiple battery-specific design objectives
\citep{cheng2026artificial}. As the complexity of the design space continues to grow,
conventional experimental, computational, and data-driven approaches alone
are no longer sufficient. A fundamentally different discovery paradigm is
therefore required \citep{wang2023scientific,lei2024materials,fuihoonnah2023generative}.

\subsection{Why generative artificial intelligence?}
\label{sec:why-genai}

Generative artificial intelligence fundamentally changes the
materials-discovery paradigm by shifting the focus from property prediction
to inverse design \citep{lei2024materials,datta2025generative,zeni2025generative,liu2023generative}. Rather than asking what properties a
given material possesses, generative models seek to construct new crystal
structures that satisfy specified design objectives. Recent advances in
diffusion models, crystal variational autoencoders, transformer
architectures, and large language models have demonstrated that AI can
generate chemically plausible crystal structures extending beyond existing
crystallographic databases, transforming AI from a screening tool into a
hypothesis-generation engine \citep{datta2025generative}.

These developments mark an important milestone in computational materials
discovery. Instead of restricting exploration to known materials or manually
enumerated substitutions, generative AI enables systematic exploration of
previously inaccessible regions of composition and crystallographic space.
For porous oxides, this capability is particularly valuable because the
diversity of possible framework architectures far exceeds what can be
explored experimentally or through conventional first-principles calculations
\citep{datta2024unlocking}. More importantly, generative AI enables researchers to search
this vast design space proactively, constructing candidate materials that
satisfy desired performance objectives rather than merely evaluating existing
ones \citep{datta2025generative}.

\subsection{Why current generative AI is still insufficient}
\label{sec:why-insufficient}

Although recent generative models have demonstrated impressive capabilities
in crystal generation, they remain largely focused on structural validity and
thermodynamic plausibility. Practical battery materials require far richer
design objectives. A generated crystal may satisfy crystallographic
constraints while lacking efficient ion transport, reversible electrochemical
behavior, mechanical durability, synthesis feasibility, or manufacturability.
Likewise, thermodynamic stability alone does not guarantee experimental
accessibility or long-term battery performance.

Overcoming these limitations requires a broader vision of AI-enabled materials
discovery. Future generative frameworks must evolve from unconstrained crystal
generation toward application-aware inverse design \citep{noh2019inverse}, where
transport functionality, electrochemistry, electro-chemo-mechanics, synthesis,
manufacturability, and uncertainty are incorporated directly into the
generative process rather than evaluated only after structure generation.
Achieving this goal will also require scientific knowledge that extends far
beyond conventional crystallographic databases, motivating the development of
autonomous knowledge-generation systems \citep{hase2019next} capable of integrating
information from the scientific literature, multiscale simulations, and
experiments.

\subsection{Scope of this perspective}
\label{sec:scope}

This perspective examines the emerging transition from database-driven
screening and unconstrained crystal generation toward physics-informed,
application-aware, and synthesis-aware autonomous discovery \citep{kwade2018current,datta2026autonomous} of
porous oxide materials for next-generation energy storage. Using porous oxides
as a representative materials platform, we examine why these materials
constitute a distinctive inverse-design problem, analyze the capabilities and
limitations of current generative AI, and propose a hierarchical
physics-informed inverse-design framework that integrates chemistry,
transport, electrochemistry, electro-chemo-mechanics, manufacturability, and
sustainability into future generative models. We further argue that realizing
this vision requires autonomous knowledge-generation systems \citep{stach2021autonomous},
ontology-driven scientific knowledge infrastructures, and synthesis-aware
closed-loop discovery \citep{kim2026closed} frameworks that tightly integrate
computational design with experimental realization.

\begin{KeyIdea}[Central thesis]
The next breakthrough in AI-enabled materials discovery will not come from
generating increasingly large numbers of nominally stable crystal structures,
but from autonomous discovery \citep{datta2026autonomous,kim2026closed} systems that integrate
scientific knowledge generation, physics-informed inverse design, multiscale
validation, synthesis planning, experimental feedback, and continual learning
within a unified scientific framework.
\end{KeyIdea}

Although our discussion focuses on porous oxide materials for energy storage,
the concepts developed here provide a general blueprint for next-generation
AI-enabled materials discovery across a broad range of functional materials
\citep{javadpour2026comprehensive}.

\section{Why Porous Oxides Are a Distinctive Inverse-Design Problem}
\label{sec:distinctive}

Porous transition-metal oxides \citep{yuan2014mixed,sun2021tailoring} occupy a unique position in
electrochemical energy storage because their performance is governed not only
by chemical composition but also by crystallographic architecture
(\cref{fig:mam-examples}b). Unlike conventional electrode materials, where
composition primarily determines electrochemical behavior, porous oxides
derive much of their functionality from interconnected tunnels, channels, and
open-framework structures that simultaneously regulate ion transport,
electronic response, structural stability, and mechanical deformation.
Consequently, designing high-performance porous oxides requires simultaneous
optimization of chemistry and framework topology, making them an unusually
challenging inverse-design problem.

\begin{figure}[tbp]
  \centering
  \includegraphics[width=0.78\linewidth]{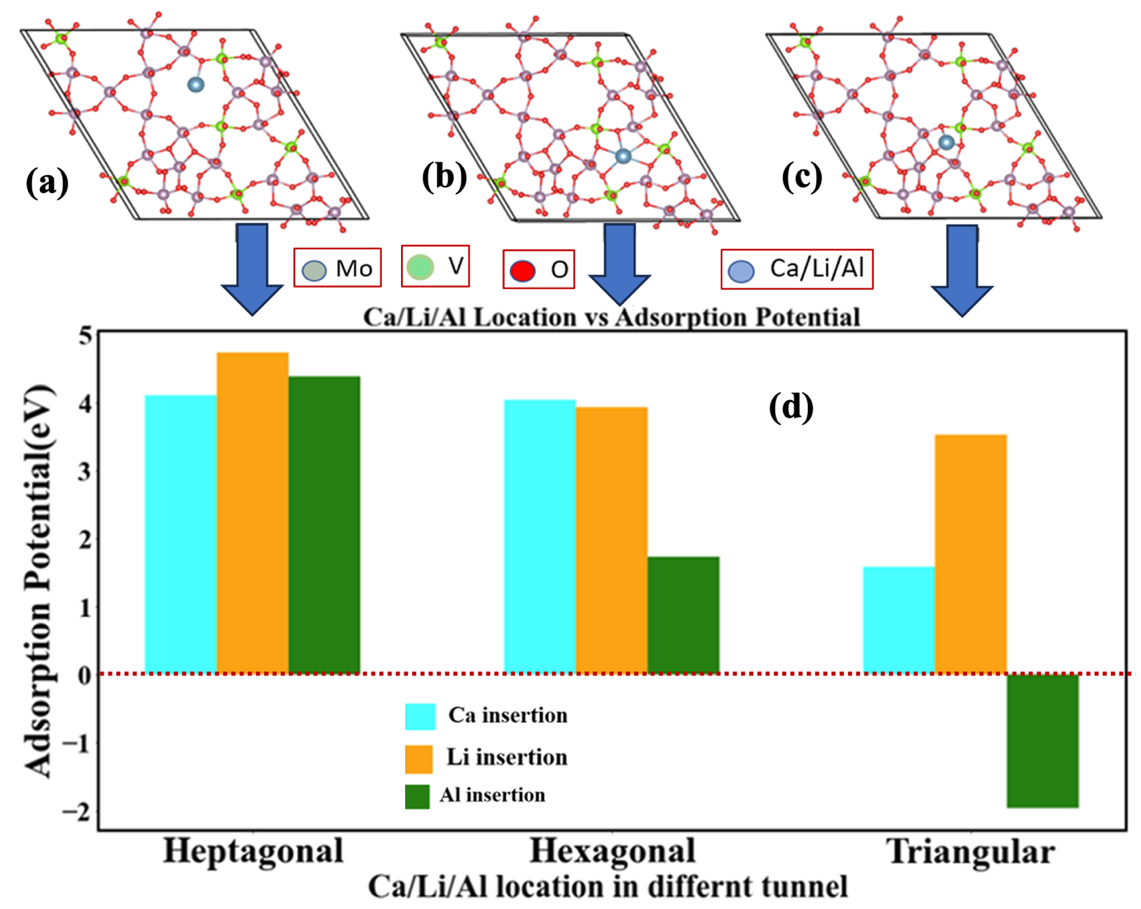}
  \caption{Comparative analysis of the adsorption potential of Ca, Li, and Al
  ions across different tunnel channels. (a)--(c) Locations of Ca, Li, and Al
  in the heptagonal, hexagonal, and triangular channels of MoVO,
  respectively. (d) Comparison of adsorption potential with respect to channel
  location. Notably, the adsorption potential remains consistent for the
  heptagonal channel, owing to its substantial channel dimensions. The
  pronounced impact of charge density becomes evident in the triangular
  channel. Reprinted (adapted) with permission from ref.~\citep{datta2024unlocking}.
  Copyright 2024 Royal Society of Chemistry.}
  \label{fig:adsorption}
\end{figure}

A defining characteristic of porous oxides is that they naturally integrate
multiple structural length scales within a single material. Conventional
microparticles remain attractive because they provide high tap density,
favorable volumetric energy density \citep{lakhnot2020aqueous}, high active-material loading,
and compatibility with scalable manufacturing. However, their long diffusion
distances often limit rate capability and promote concentration gradients,
stress accumulation, and structural degradation during cycling.
Nanostructured electrodes alleviate many of these limitations by shortening
diffusion pathways, but their high surface area frequently leads to excessive
surface reactions, low tap density, particle aggregation, and increased
manufacturing complexity. Porous oxides provide a fundamentally different
strategy by retaining microscale particles while incorporating
crystallographically defined nanoscale transport channels, thereby combining
the manufacturing advantages of microparticles with the transport
characteristics of nanostructured electrodes. In this sense, porous oxides
function as natural multiscale active materials.

Their advantages extend beyond rapid ion transport. Interconnected framework
networks provide multiple insertion sites, shorten effective diffusion
distances, and partially accommodate insertion-induced lattice distortion
during cycling, thereby improving rate capability and structural durability.
However, these benefits are not guaranteed by porosity alone. Excessive pore
volume reduces framework density and volumetric energy density, poorly
connected channels contribute little to long-range transport, and highly open
structures may become mechanically fragile or electrochemically unstable.
Successful porous oxide design therefore requires balancing transport
functionality, volumetric energy density, electrochemical performance, and
mechanical stability rather than maximizing any single descriptor.

\begin{figure}[tbp]
  \centering
  \includegraphics[width=0.66\linewidth]{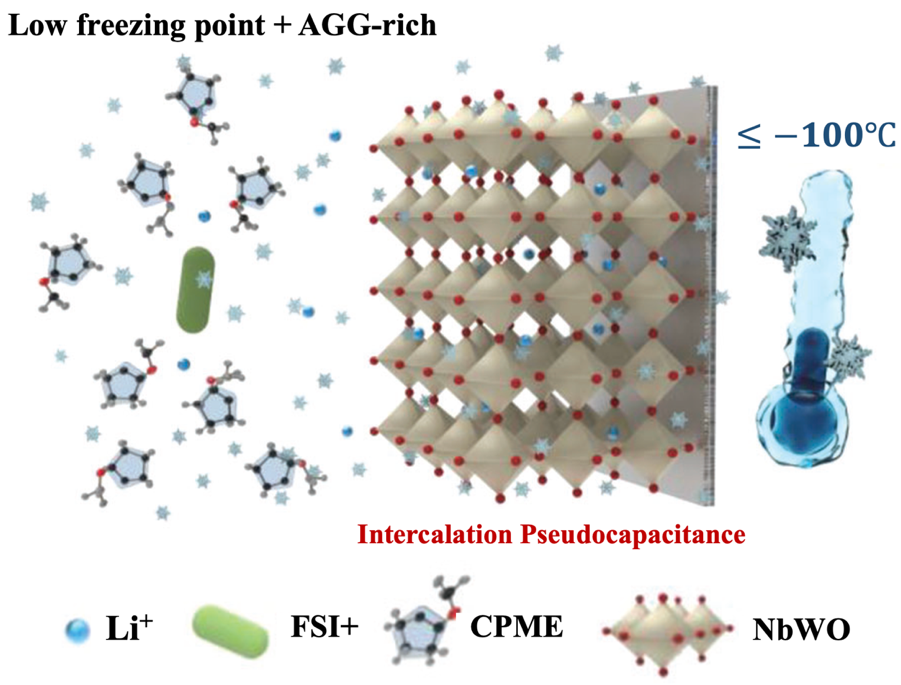}
  \caption{NbWO electrode and CPME-based electrolyte properties enabling
  extreme low-temperature ($\leq -100\,^{\circ}$C) battery cycling. Reprinted
  (adapted) with permission from ref.~\citep{kim2024enabling}. Copyright 2024 Wiley.}
  \label{fig:lowT}
\end{figure}

These coupled design requirements become even more demanding for
multivalent-ion batteries (\cref{fig:adsorption}) based on \ion{Mg}{2+},
\ion{Ca}{2+}, \ion{Zn}{2+}, \ion{Al}{3+}, etc. Compared with \ion{Li}{+},
multivalent ions exhibit stronger electrostatic interactions with both the
electrolyte and host framework, producing larger desolvation barriers, slower
diffusion, and greater insertion-induced lattice distortion. Consequently,
pore architecture becomes an explicit design variable rather than merely a
structural feature. Channel dimensions, bottleneck geometry, connectivity, and
local coordination environments must all be tailored to the intended charge
carrier. The same crystallographic features that facilitate multivalent-ion
transport also make porous oxides attractive for fast-charging and
low-temperature batteries \citep{kim2024enabling} (\cref{fig:lowT}), where interconnected
transport pathways help mitigate kinetic limitations. Porous oxides should
therefore be viewed as system-level materials whose performance emerges from
coupled interactions among crystal structure, transport, electrochemistry,
mechanics, and the surrounding electrochemical environment.

The remarkable structural diversity of porous oxide materials further expands
this design space. Wadsley--Roth crystallographic shear phases,
molybdenum--vanadium oxides, manganese oxides, and related open-framework
materials illustrate how variations in channel geometry, framework topology,
defect chemistry, and compositional substitution produce markedly different
transport pathways, insertion mechanisms, and electrochemical behavior. Rather
than representing a single class of materials, porous oxides comprise a
diverse family of crystallographic frameworks whose functionality is
intrinsically linked to structural topology.

This diversity ultimately explains why porous oxides present a fundamentally
different inverse-design problem from conventional inorganic materials. The
objective is not simply to identify a chemically stable crystal, but to
discover a framework in which chemistry, pore architecture, transport,
electrochemistry, electro-chemo-mechanics \citep{datta2024ecm}, and synthesis remain
simultaneously optimized. These requirements are strongly coupled and often
competing, making the optimal material a carefully balanced compromise rather
than the maximization of any individual property.

\begin{KeyTakeaway}
For generative artificial intelligence, this distinction has profound
implications. Current crystal-generation models primarily optimize structural
validity and thermodynamic plausibility. Successful porous oxide discovery,
however, requires models capable of reasoning simultaneously about
crystallography, ion-specific transport, electrochemistry, mechanics, and
synthesis feasibility. The target of discovery is therefore not merely a
stable crystal, but a functional porous architecture engineered for practical
electrochemical energy storage.
\end{KeyTakeaway}

This distinction motivates the transition from generic crystal generation to
the physics-informed, application-aware inverse-design framework developed in
the following sections.

\section{Why Conventional Discovery Is No Longer Sufficient}
\label{sec:conventional}

The discovery of next-generation porous oxide materials is constrained not by
the absence of promising structural concepts but by the enormous complexity of
the underlying design space. Unlike conventional electrode materials, porous
oxides must simultaneously satisfy requirements involving chemical
composition, crystallographic topology, pore architecture, ion transport,
electrochemical functionality, mechanical durability, and synthesis
feasibility. These characteristics are strongly coupled, creating a
multidimensional optimization problem \citep{jin2008pareto} in which improving one
property often compromises another. Consequently, discovering
high-performance porous oxides requires exploring a vast landscape of
chemically plausible compositions and framework architectures rather than
optimizing a single material descriptor.

This complexity arises from numerous interacting design variables.
Transition-metal oxides exhibit diverse compositions, oxidation states,
polymorphs, defect configurations, and framework topologies, while small
compositional changes can alter tunnel geometry, electronic structure,
diffusion pathways, and phase stability. Different charge carriers further
modify transport behavior through differences in size, charge density, and
solvation, and operating conditions such as temperature, electrolyte
composition, and state of charge introduce additional constraints. Even after
restricting the search to chemically meaningful candidates, the remaining
design space remains far beyond what can be explored exhaustively.

Experimental discovery remains the definitive route for validating new
materials but is inherently slow and resource intensive. Every candidate
requires synthesis optimization, structural characterization, electrode
fabrication, and electrochemical testing, while small variations in precursor
chemistry or processing conditions can produce different polymorphs and
competing phases. First-principles calculations \citep{datta2024unlocking,sharma2018amorphous} provide
atomic-scale insight into stability, ion transport, and electrochemical
behavior, yet they become computationally prohibitive for porous
transition-metal oxides containing large unit cells, crystallographic
disorder, multiple insertion sites, and competing phases. Consequently, both
experiments and high-fidelity simulations are best suited for validating
carefully selected candidates rather than systematically exploring the entire
design space.

Molecular dynamics (MD) simulations \citep{kim2014atomistic} are limited by the lack of
suitable interatomic potentials \citep{oluwajobi2011effect} for many materials, including
porous oxides.
Machine-learning interatomic potentials have found significant applications in
various materials science problems \citep{chan2019machine,deringer2019machine}, but this approach also
needs significant first-principles training data \citep{chan2019machine,wang2024machine}.
Machine-learning surrogate models have substantially accelerated
computational screening by replacing selected first-principles calculations
with inexpensive property predictions \citep{sharma2022developing}. However, these approaches
remain fundamentally evaluative: they estimate the properties of known or
manually generated materials rather than discovering entirely new crystal
structures optimized for specific battery applications. This limitation is
particularly important for porous oxides because the most promising framework
architectures may lie well beyond currently known material families.

Generative artificial intelligence fundamentally changes this paradigm by
shifting the central question from ``What are the properties of this
material?'' to ``What material should be designed to achieve these
properties?'' \citep{datta2025generative}. By learning statistical representations of crystal
chemistry, diffusion models, crystal variational autoencoders, transformer
architectures, and large language models can propose previously unexplored
crystal structures, dramatically expanding the searchable design space.
However, crystal generation alone does not constitute materials discovery.
Generated structures remain computational hypotheses that require
progressively higher levels of validation through machine-learning screening,
first-principles calculations, multiscale simulations, and ultimately
experimental synthesis and characterization.

The central limitation of conventional discovery therefore lies not in any
individual methodology but in the absence of an integrated discovery
strategy. Experiments, first-principles calculations, machine learning, and
generative AI each contribute distinct capabilities, yet none is sufficient
independently. Future porous oxide discovery should instead combine these
complementary approaches within a physics-informed inverse-design framework,
where generative AI proposes candidate materials, predictive models perform
rapid multi-objective evaluation, high-fidelity simulations establish
mechanistic credibility, and experiments provide definitive validation.
Rather than exhaustively searching the design space, such a framework directs
exploration toward materials where chemical plausibility, transport
functionality, electrochemical performance, mechanical durability, and
synthesis feasibility converge.

The following section illustrates this transition through a representative
case study of AI-driven porous oxide discovery. Rather than emphasizing
algorithmic performance alone, the case study highlights the key scientific
lessons that motivate the physics-informed, application-aware discovery
framework developed throughout the remainder of this perspective.

\section{Case Study: What First-Generation Generative AI Discovery Revealed}
\label{sec:casestudy}

\begin{figure}[tbp]
  \centering
  \includegraphics[width=\figw]{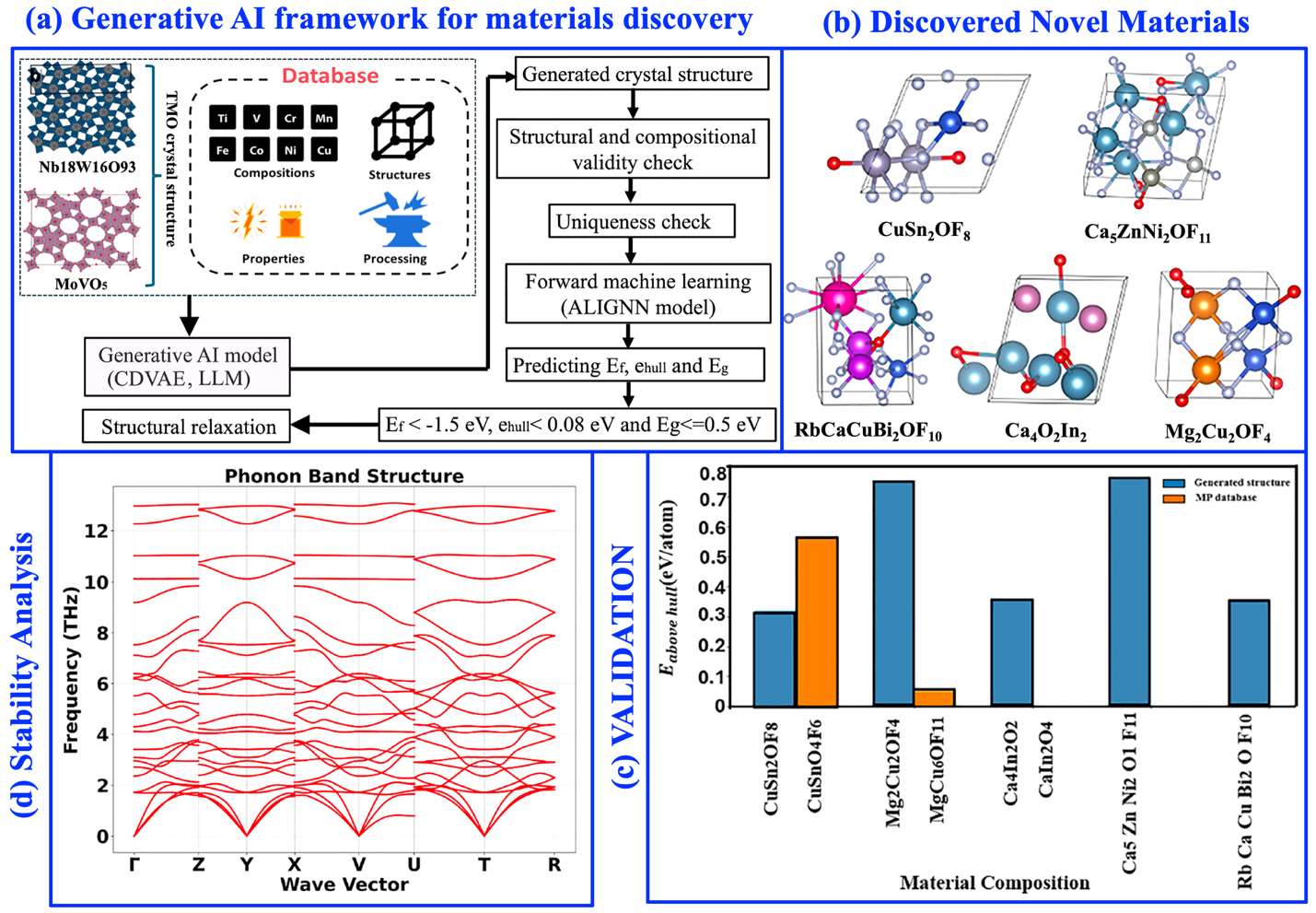}
  \caption{Case study: generative AI for materials discovery. (a) Overview of
  the generative AI framework, trained on the dataset from the Materials
  Project database, for constructing crystal structures. (b) Overview of five
  transition-metal oxide (TMO)-based structures generated by the CDVAE model,
  highlighting large open-tunnel frameworks. (c) Comparison of $\Ehull$ for
  generated TMO structures and Materials Project database entries
  \citep{jain2013materials}. (d) Phonon dispersion analysis of Ca$_4$In$_2$O$_2$, providing
  a thorough examination of its dynamical stability. Reprinted (adapted) with
  permission from ref.~\citep{datta2025generative}. Copyright 2025 Cell Press.}
  \label{fig:casestudy}
\end{figure}

\begin{figure}[tbp]
  \centering
  \includegraphics[width=\figw]{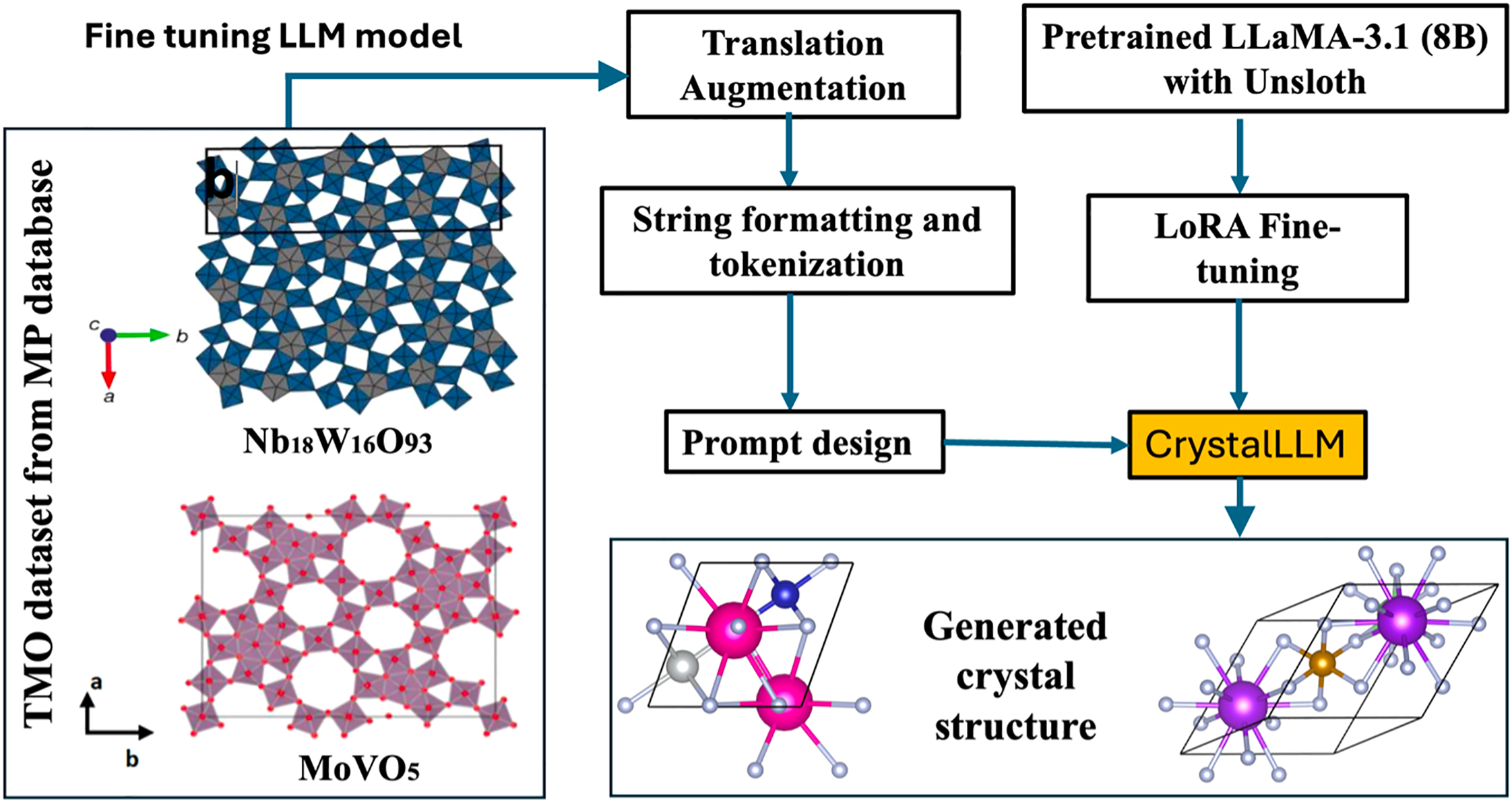}
  \caption{Overview of fine-tuning the LLaMA~3.1 (8B) model for generating
  crystal structures. Low-rank adaptation (LoRA) trains only a subset of the
  model parameters (the low-rank matrices) to adapt the pre-trained model to
  the crystal-generation task. Reprinted (adapted) with permission from
  ref.~\citep{datta2025generative}. Copyright 2025 Cell Press.}
  \label{fig:llm}
\end{figure}

\begin{figure}[tbp]
  \centering
  \includegraphics[width=\figw]{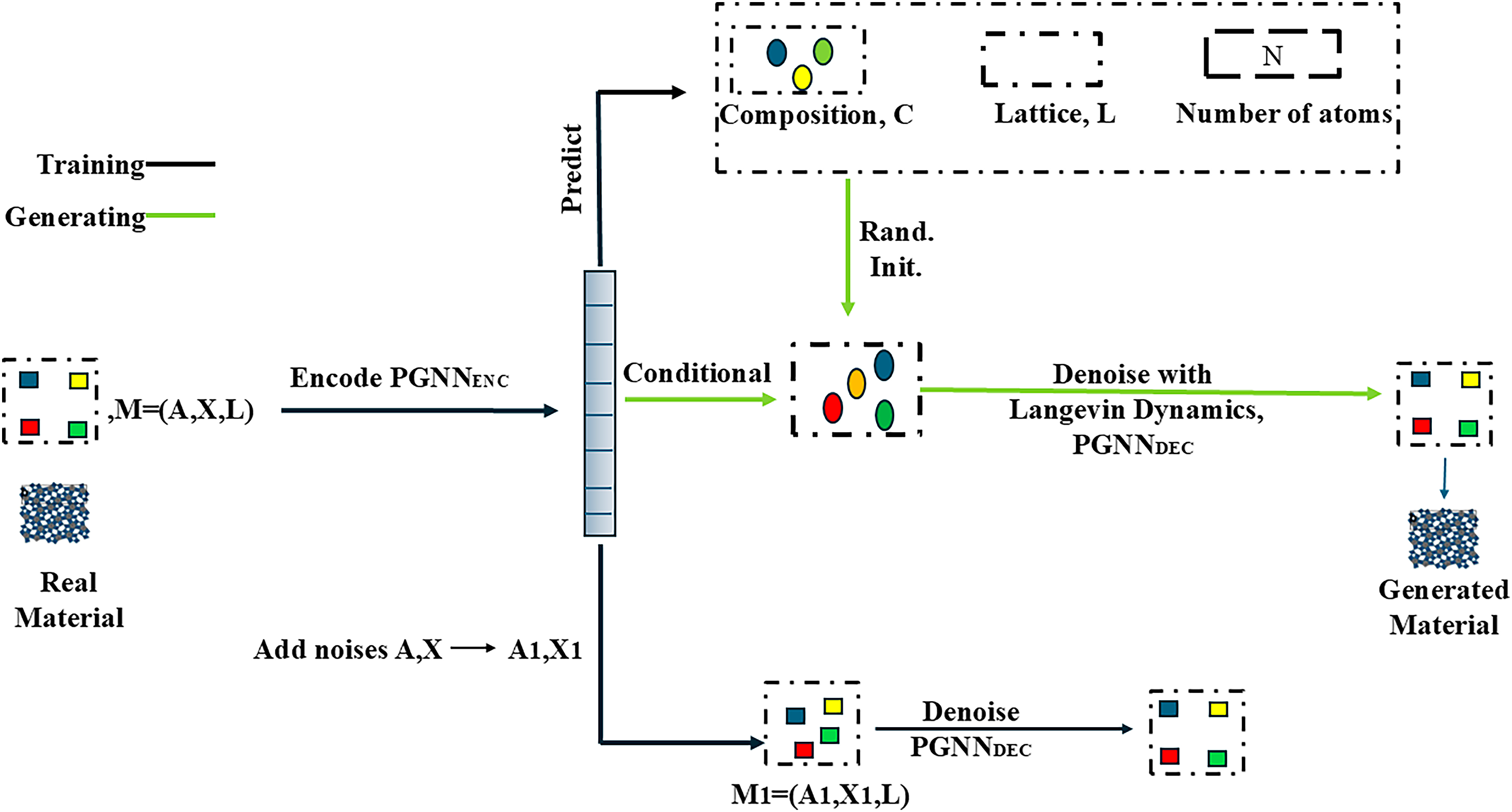}
  \caption{Architecture of the CDVAE model for generating stable material
  structures. The model combines an SE(3)-equivariant periodic graph neural
  network (PGNN) encoder, an attribute predictor, and an SE(3)-equivariant
  decoder using Langevin dynamics. The encoder maps atomic types,
  coordinates, and lattice information into a latent space, from which the
  attribute predictor estimates composition, lattice parameters, and atom
  count. The decoder refines noisy latent representations through iterative
  denoising to generate physically stable materials, ensuring that periodic
  and symmetry properties are maintained. Reprinted (adapted) with permission
  from ref.~\citep{datta2025generative}. Copyright 2025 Cell Press.}
  \label{fig:cdvae}
\end{figure}

The transition from conventional materials screening to generative inverse
design is illustrated by a recent dual-model framework \citep{datta2025generative} for porous
oxide discovery (\cref{fig:casestudy}) that combined a fine-tuned large
language model (LLM, \cref{fig:llm}) and a crystal diffusion variational
autoencoder (CDVAE, \cref{fig:cdvae}), followed by hierarchical validation
using structural verification, machine-learning property prediction, density
functional theory (DFT), and thermodynamic analysis. More important than the
individual computational results, this study revealed several broader lessons
regarding the capabilities and limitations of first-generation generative AI
for materials discovery \citep{datta2025generative}.

\begin{KeyFinding}[Lesson 1]
\Lesson{Different generative models explore materials space differently.}
The diffusion model (\cref{fig:cdvae}) preferentially explored broader regions
of crystallographic space, generating structurally diverse candidates that
often required substantial relaxation before approaching local energy minima
\citep{datta2025generative,kim2024diffusion}. In contrast, the language model (\cref{fig:llm}) remained
closer to known crystallographic distributions and produced a larger fraction
of structures near thermodynamic equilibrium after DFT relaxation. Rather than
identifying one architecture as universally superior, the study revealed an
inherent exploration--exploitation trade-off. Future discovery systems will
likely combine these complementary capabilities rather than relying on a
single generative architecture.
\end{KeyFinding}

\begin{KeyFinding}[Lesson 2]
\Lesson{Crystal generation is only the beginning.}
Although generative models can rapidly produce thousands of candidate
structures, only a small fraction survive progressively more demanding
validation (\cref{fig:casestudy}). Structural plausibility, charge
neutrality, novelty, machine-learning screening, DFT relaxation,
thermodynamic stability, and ultimately electrochemical functionality form a
hierarchy of increasingly stringent filters. The scientific value of
generative AI therefore lies not in the number of generated crystals but in
efficiently prioritizing promising candidates for higher-fidelity computation
and experimental validation. Materials discovery should consequently be viewed
as a hierarchical validation process rather than a crystal-generation
exercise.
\end{KeyFinding}

\begin{KeyFinding}[Lesson 3]
\Lesson{Statistical learning does not guarantee application-aware design.}
An unexpected outcome of the study was that many generated structures no
longer belonged to the intended class of porous oxide materials despite
training on oxide-rich datasets. Both the diffusion model and the language
model frequently generated chemically plausible compounds lacking oxygen
altogether. This observation highlights an important limitation of current
generative AI: learning the statistical distribution of crystal structures
does not necessarily imply learning the scientific objective of a
materials-design problem. Future models must therefore incorporate explicit
compositional constraints, oxidation-state consistency, pore-architecture
objectives, and chemistry-aware decoding to align generation with the
intended application.
\end{KeyFinding}

\begin{KeyFinding}[Lesson 4]
\Lesson{Thermodynamic stability is not battery performance.}
Current generative workflows primarily evaluate formation energy, energy above
the convex hull, and electronic properties because these quantities provide
efficient measures of structural plausibility. However, thermodynamic
stability alone does not guarantee ion accessibility, favorable migration
pathways, reversible electrochemistry, mechanical durability, or synthesis
feasibility. Conversely, metastable materials may prove experimentally
accessible through kinetic stabilization. This distinction emphasizes that
crystal discovery and battery-material discovery are fundamentally different
objectives.
\end{KeyFinding}

\begin{KeyFinding}[Lesson 5]
\Lesson{Computational predictions remain scientific hypotheses.}
As generative AI becomes increasingly capable of proposing previously unknown
crystal structures, favorable computational metrics should not be equated
with materials discovery. DFT relaxation, convex-hull analysis, and phonon
calculations progressively strengthen computational confidence, but they do
not establish synthesizability, electrochemical performance, or technological
viability. Computationally generated materials should therefore be viewed as
scientifically informed hypotheses requiring experimental validation rather
than confirmed discoveries.
\end{KeyFinding}

Collectively, these lessons extend well beyond the specific computational
workflow considered here. First-generation generative AI has demonstrated that
autonomous crystal generation is feasible, but it has also clarified the
limitations that must be overcome before AI can routinely discover deployable
battery materials. Future systems must evolve from unconstrained crystal
generation toward physics-informed, application-aware inverse design, where
chemistry, crystallography, transport, electrochemistry, mechanics,
synthesis, and uncertainty are incorporated directly into the generative
process. These lessons motivate the hierarchical inverse-design framework
developed in the following section.

\section{From Generic Generation to Physics-Informed Porous-Oxide Design}
\label{sec:tiers}

The next generation of autonomous materials discovery (\cref{fig:framework})
should transition from generic crystal generation to physics-informed inverse
design \citep{karniadakis2021physics}, where scientific knowledge is embedded directly into the
generative process rather than imposed through post-generation filtering.
Instead of asking whether a generated crystal satisfies a sequence of
independent screening criteria, future AI systems should learn to generate
materials that already satisfy the coupled requirements governing practical
electrochemical performance.

This perspective organizes these requirements into a hierarchical framework
consisting of seven progressively integrated design tiers
(\cref{sch:tiers}). Each tier introduces an additional level of physical
realism while building upon the constraints established by the preceding
tier. Together, they provide a roadmap for transforming generative AI from a
crystal-generation engine into an autonomous battery-materials design
framework.

\begin{scheme}[!htb]
  \centering
  \begin{tikzpicture}[
      tier/.style={ddnode, text width=88mm, minimum height=9.5mm,
                   align=left, font=\footnotesize},
      node distance=2.2mm]
    \node[ddstructure, tier, fill=DDNavy!9, draw=DDNavy!60] (t1)
      {\textbf{Tier 1\quad Chemical and crystallographic validity}\\
       \textit{Generate chemically credible crystal structures}};
    \node[ddstructure, tier, fill=DDNavy!9, draw=DDNavy!60, above=of t1] (t2)
      {\textbf{Tier 2\quad Thermodynamic and dynamical viability}\\
       \textit{Generate physically viable materials}};
    \node[ddmethod, tier, fill=DDIndigo!9, draw=DDIndigo!60, above=of t2] (t3)
      {\textbf{Tier 3\quad Porosity and transport functionality}\\
       \textit{Generate functional transport networks}};
    \node[ddmethod, tier, fill=DDIndigo!9, draw=DDIndigo!60, above=of t3] (t4)
      {\textbf{Tier 4\quad Electrochemical functionality}\\
       \textit{Generate electrochemically active electrodes}};
    \node[ddmethod, tier, fill=DDIndigo!9, draw=DDIndigo!60, above=of t4] (t5)
      {\textbf{Tier 5\quad Electro-chemo-mechanical durability}\\
       \textit{Generate durable materials}};
    \node[ddresult, tier, fill=DDTeal!11, draw=DDTeal!70, above=of t5] (t6)
      {\textbf{Tier 6\quad Electrode and cell compatibility}\\
       \textit{Generate cell-compatible electrodes}};
    \node[ddresult, tier, fill=DDTeal!11, draw=DDTeal!70, above=of t6] (t7)
      {\textbf{Tier 7\quad Manufacturability and sustainability}\\
       \textit{Generate deployable technologies}};
    \draw[ddflow, line width=1.1pt]
      ([xshift=-6mm]t1.south west) -- ([xshift=-6mm]t7.north west)
      node[midway, sloped, above, font=\scriptsize\color{DDMeta}]
      {increasing physical realism};
    \draw[decorate, decoration={brace, amplitude=4pt}, draw=DDNavy!70]
      ([xshift=3mm]t5.north east) -- ([xshift=3mm]t1.south east)
      node[midway, right=5pt, font=\scriptsize, text=DDNavy, align=left]
      {intrinsic properties\\of the active material};
    \draw[decorate, decoration={brace, amplitude=4pt}, draw=DDTeal!80]
      ([xshift=3mm]t7.north east) -- ([xshift=3mm]t6.south east)
      node[midway, right=5pt, font=\scriptsize, text=DDTeal!90!black, align=left]
      {battery system\\and deployment};
  \end{tikzpicture}
  \caption{The seven-tier physics-informed inverse-design hierarchy proposed in
  \cref{sec:tiers}. Each tier adds a level of physical realism to the
  constraints of the tiers below it, progressing from a chemically credible
  crystal to a deployable energy-storage technology. Tiers~1--5 concern the
  intrinsic properties of the active material; Tiers~6 and~7 concern its
  function within a complete cell and its deployment at scale.}
  \label{sch:tiers}
\end{scheme}

\subsection{Tier 1: Chemical and crystallographic validity}
\label{sec:tier1}
\TierGoal{Generate chemically credible crystal structures.}

The foundation of autonomous materials discovery is the generation of
chemically and crystallographically credible crystal structures. Regardless of
the sophistication of subsequent property prediction or first-principles
validation, structures that violate fundamental chemical principles cannot
become practical battery materials. First-generation generative models
typically enforce these constraints through post-generation filtering,
discarding chemically unrealistic candidates after they have already been
generated. Future inverse-design frameworks should instead incorporate these
constraints directly into the generation process so that every candidate
begins as a chemically credible hypothesis.

This requires satisfying several fundamental descriptors simultaneously.
Generated structures should exhibit chemically meaningful stoichiometry,
charge neutrality, oxidation states consistent with realistic
transition-metal chemistry, physically reasonable coordination environments,
and crystallographic periodicity compatible with stable crystalline solids. At
the same time, they should remain genuinely novel without drifting into
chemically implausible regions of structural space. Novelty should therefore
arise from scientifically meaningful extrapolation of known crystal chemistry
rather than unconstrained statistical sampling.

Collectively, these descriptors define the chemically accessible design space
within which autonomous discovery should operate. Embedding them directly into
generative models not only reduces the number of unrealistic candidates but
also shifts AI from learning statistical regularities toward learning the
underlying principles of inorganic chemistry. Chemical and crystallographic
validity therefore establishes the foundation of the proposed inverse-design
hierarchy, providing the minimum scientific credibility required before
higher-level optimization of thermodynamic stability, transport,
electrochemistry, durability, and manufacturability can begin.

\subsection{Tier 2: Thermodynamic and dynamical viability}
\label{sec:tier2}
\TierGoal{Generate physically viable materials.}

Chemical validity (Tier~1) alone does not guarantee that a generated crystal
can exist as a viable material. A composition may satisfy stoichiometric and
crystallographic constraints yet remain energetically unstable or transform
into competing phases. The second tier of physics-informed inverse design
therefore evaluates whether a chemically credible crystal is also
thermodynamically and dynamically viable under realistic synthesis and
operating conditions.

The first level of assessment is equilibrium thermodynamics. Formation energy
measures the energetic favorability of the proposed framework, while the
energy above the convex hull evaluates its stability relative to competing
phases (\cref{fig:casestudy}). Dynamic stability (\cref{fig:casestudy})
provides complementary information by determining whether the relaxed crystal
occupies a mechanically stable configuration on the potential-energy surface.
Because a material may be thermodynamically metastable yet dynamically
stable, or vice versa, both aspects are required to establish physical
viability.

However, practical battery materials rarely exist under equilibrium
conditions. Synthesis commonly involves rapid heating, quenching, vapor
deposition, or electrochemical formation, while battery operation continuously
changes ion concentration, chemical potential, and temperature.
Finite-temperature effects, including vibrational and configurational
entropy, magnetic contributions, and thermal expansion, can therefore
significantly modify phase stability. Likewise, many technologically important
materials remain experimentally accessible despite being metastable because
kinetic barriers suppress transformation into lower-energy phases. Future
inverse-design frameworks should therefore move beyond zero-temperature
equilibrium calculations to estimate finite-temperature stability, kinetic
persistence, and experimentally accessible metastable states.

Real materials also contain vacancies, disorder, grain boundaries, and other
defects that influence both stability and electrochemical behavior. Rather
than evaluating ideal defect-free crystals alone, future AI systems should
progressively learn the stability of physically realizable structural
ensembles.

The objective of this tier is therefore not simply to identify the
lowest-energy crystal, but to determine whether a generated framework can
realistically survive synthesis and subsequent electrochemical operation. By
incorporating equilibrium thermodynamics, lattice dynamics, finite-temperature
behavior, metastability, and defect tolerance directly into the generative
process, AI can evolve from predicting energetically favorable structures to
designing physically realizable materials.

\subsection{Tier 3: Porosity and transport functionality}
\label{sec:tier3}
\TierGoal{Generate functional transport networks.}

Thermodynamic viability (Tier~2) establishes whether a crystal can exist, but
it does not determine whether it can function as an effective battery
electrode. For porous oxides, electrochemical performance depends not on the
presence of internal void space alone, but on interconnected pathways that
enable rapid and reversible ion transport. The objective of the third tier is
therefore to determine whether a generated framework possesses
\emph{functional} porosity rather than simply \emph{geometric} porosity.

This distinction is fundamental. A crystallographically open framework is not
necessarily electrochemically accessible. Large cavities may remain
ineffective if they are isolated by narrow bottlenecks, disconnected channels,
or unfavorable electrostatic environments, whereas relatively modest pore
volumes can exhibit excellent performance when transport pathways are
continuous, energetically favorable, and matched to the intended charge
carrier. Future generative models must therefore distinguish geometric
porosity from ion-accessible porosity, since only the latter governs practical
battery performance.

Transport functionality is determined by several coupled descriptors.
Accessible pore diameter controls ion entry, bottleneck size governs
long-range migration, and transport dimensionality, pore connectivity, and
tortuosity determine diffusion efficiency throughout the framework. These
descriptors are inherently ion dependent. A channel that readily accommodates
\ion{Li}{+} may become ineffective for \ion{Mg}{2+} or \ion{Al}{3+} because of
their larger effective size, stronger electrostatic interactions, and
different coordination preferences. Consequently, pore architecture must
always be optimized for the intended charge carrier rather than evaluated
using generic geometric metrics.

Geometry alone, however, cannot fully describe ion transport. Migration
barriers, local coordination environments, electrostatic interactions, and
diffusion bottlenecks collectively determine transport kinetics, rate
capability, and low-temperature performance (\cref{fig:lowT}). Moreover, ion
transport begins before ions enter the crystal. Desolvation and interfacial
charge transfer at the electrode--electrolyte interface \citep{cresce2014situ,li2023review} often
become dominant kinetic barriers, particularly for multivalent-ion batteries
(\cref{fig:adsorption}). Functional transport therefore depends on the coupled
effects of crystallographic topology, local chemistry, and interfacial
processes rather than pore geometry alone.

Current generative AI largely treats porosity as a structural descriptor
evaluated after crystal generation \citep{datta2025generative}. Future inverse-design
frameworks should instead optimize transport functionality directly during
generation by learning the coupled relationships among pore topology, channel
connectivity, migration barriers, ion-specific accessibility, and interfacial
transport. The objective is no longer to generate crystals that merely contain
pores, but to design frameworks that admit, transport, and reversibly release
the intended charge carrier while maintaining structural integrity under
realistic operating conditions.

\subsection{Tier 4: Electrochemical functionality}
\label{sec:tier4}
\TierGoal{Generate electrochemically active electrodes.}

Chemical validity (Tier~1), thermodynamic viability (Tier~2), and functional
ion transport (Tier~3) establish the structural foundation of a porous oxide,
but they do not determine whether it will function as an effective battery
electrode. Electrochemical performance emerges only when the host framework can
repeatedly store and release ions while maintaining favorable redox chemistry,
efficient charge transport, and structural integrity. The objective of the
fourth tier is therefore to optimize electrochemical functionality rather than
crystallographic stability alone.

Electrochemical functionality is inherently multidimensional. Operating
voltage depends on the coupled effects of crystal structure, transition-metal
chemistry, local coordination, and guest-ion interactions, while capacity
depends not only on the number of redox-active sites but also on their
accessibility throughout repeated cycling. Practical battery design must also
balance gravimetric and volumetric energy density. Highly porous frameworks
may achieve excellent gravimetric capacity at the expense of volumetric
performance, whereas dense structures often maximize volumetric energy density
while limiting ion transport. Future inverse-design frameworks must therefore
optimize these competing objectives simultaneously rather than maximizing any
single metric.

Electrochemical performance also evolves continuously during battery
operation. Efficient energy storage requires rapid transport of both ions and
electrons together with favorable interfacial charge-transfer kinetics.
Electronic conductivity, ionic conductivity, redox accessibility, and phase
stability all change with ion concentration, oxidation state, and local
bonding environment. Many electrode materials undergo concentration-dependent
phase evolution, including solid-solution behavior, intermediate phases, and
reversible phase transformations that directly influence voltage, transport,
and capacity. Future AI models should therefore treat electrochemical behavior
as a dynamic process rather than a collection of static material properties.

Ultimately, the defining characteristic of a practical battery material is
reversibility. High theoretical capacity has little value if repeated cycling
progressively degrades transport pathways, electrochemical activity, or
structural integrity through irreversible phase transformations,
transition-metal dissolution, electrolyte side reactions, or pore collapse.
Reversibility should therefore become an explicit design objective rather than
an indirect consequence of equilibrium stability.

Current generative AI typically evaluates electrochemical properties only
after crystal generation. Future inverse-design frameworks should instead
incorporate electrochemical objectives directly into the generative process,
enabling AI to generate frameworks that already satisfy target voltage
windows, balanced gravimetric and volumetric energy density, efficient ionic
and electronic transport, accessible redox centers, and reversible cycling.
The objective is no longer to generate stable porous crystals, but to design
electrochemically functional materials optimized for practical energy storage.

\subsection{Tier 5: Electro-chemo-mechanical durability}
\label{sec:tier5}
\TierGoal{Generate durable materials.}

\begin{figure}[tbp]
  \centering
  \includegraphics[width=\figw]{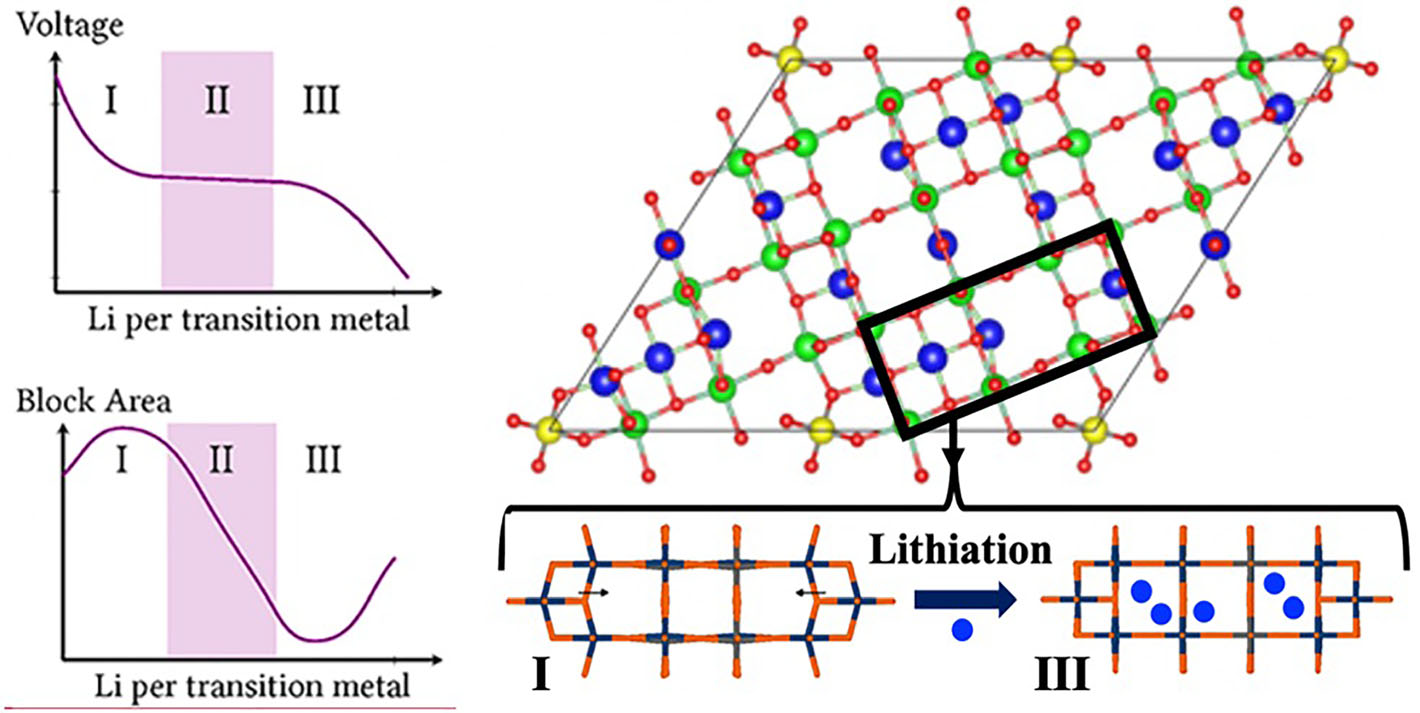}
  \caption{Structural change of the Wadsley--Roth crystallographic block of
  Li$_x$Nb$_{12}$WO$_{33}$ upon lithiation (DFT study). Reprinted (adapted)
  with permission from ref.~\citep{yang2021wadsley}. Copyright 2021 Wiley.}
  \label{fig:wadsley}
\end{figure}

Thermodynamic stability (Tier~2) and electrochemical functionality (Tier~4)
determine
whether a porous oxide can store electrical energy, but they do not determine
whether it can continue to do so over hundreds or thousands of
charge--discharge cycles. During battery operation, ion insertion continuously
modifies the material's composition, electronic structure, lattice geometry,
and internal stress state (\cref{fig:wadsley}). The objective of the fifth
tier is therefore to optimize electro-chemo-mechanical durability, ensuring
that the material remains functional throughout its operational lifetime
rather than during only its initial cycle.

Unlike conventional structural materials, battery electrodes operate under
continuous nonequilibrium chemical loading. Spatially nonuniform ion
insertion generates concentration gradients that produce chemical strain and
diffusion-induced stress even in the absence of external mechanical loading.
Because the resulting stress field depends strongly on crystallographic
architecture, transport pathways and mechanical stability are inherently
coupled. Open-framework materials such as Wadsley--Roth oxides illustrate this
behavior particularly well: crystallographic shear planes can partially
accommodate insertion-induced deformation, while the same transport channels
that facilitate rapid ion migration may also become sites of stress
concentration, crack initiation, or structural collapse (\cref{fig:wadsley}).

Mechanical degradation arises from the accumulation of these coupled processes
during repeated cycling. Local lattice distortion may evolve into irreversible
phase transformation, amorphization, tunnel collapse, particle fracture, or
loss of electrical connectivity, progressively reducing accessible ion
pathways and reversible capacity. A material's resistance to these
degradation mechanisms depends not only on intrinsic mechanical
properties (including elastic stiffness, fracture toughness, and structural
compliance) but also on the strong two-way coupling between stress and
diffusion. Mechanical stress alters ion migration and local chemical
potential, while nonuniform ion transport continuously generates new stress
\citep{yang2021wadsley}. Future AI models should therefore treat transport and mechanics
as coupled state variables rather than independent material descriptors.

\begin{figure}[tbp]
  \centering
  \includegraphics[width=0.74\linewidth]{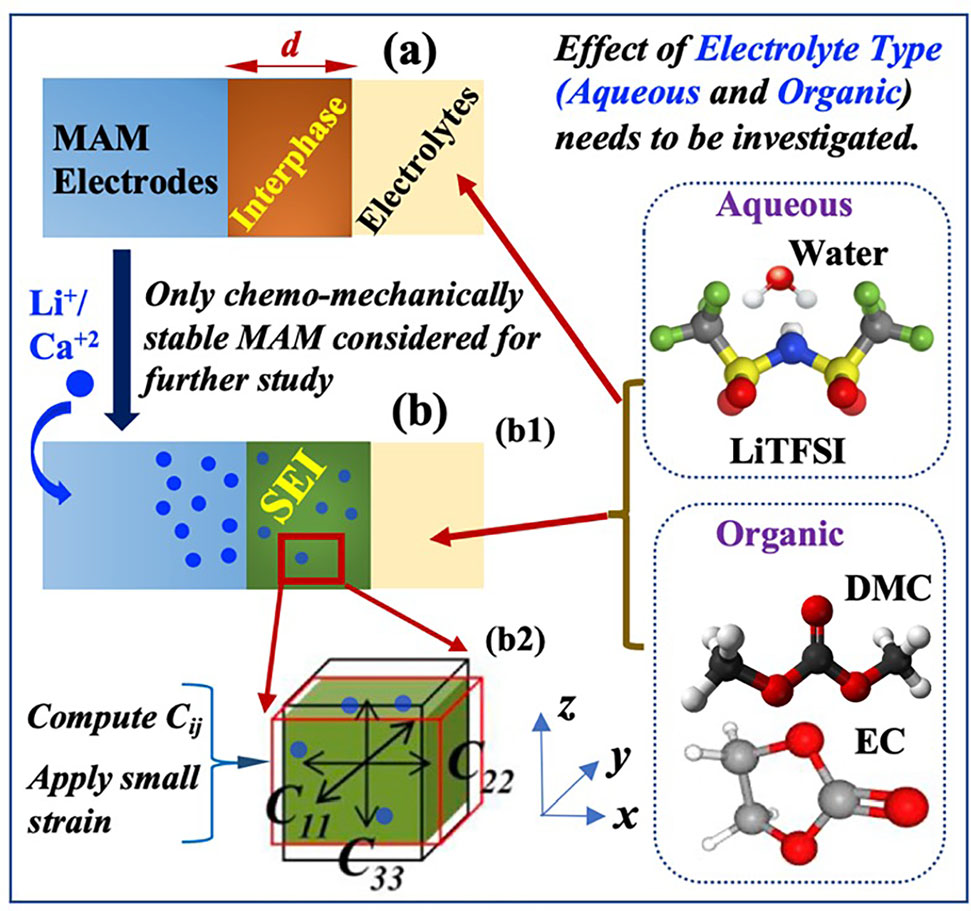}
  \caption{(a) Interphase formation between a multiscale active material (MAM)
  electrode and the electrolyte. (b) SEI formation (b1) and SEI mechanical
  properties \citep{kim2011effect} (b2). Reprinted (adapted) with permission from
  ref.~\citep{datta2024ecm}. Copyright 2024 The Minerals, Metals \& Materials Society.}
  \label{fig:sei}
\end{figure}

Durability also depends on factors extending beyond the crystal itself.
Temperature influences diffusion kinetics \citep{lu1997diffusion}, stress relaxation, and
phase stability, while repeated interaction with the electrolyte governs the
formation and evolution of the solid electrolyte interphase \citep{li2023review} (SEI,
\cref{fig:sei}), whose mechanical integrity strongly affects long-term cycling
stability. Likewise, interfacial adhesion among the active material, binder,
and current collector must remain intact despite repeated expansion and
contraction. These coupled interfacial processes remain among the least
explored aspects of AI-guided battery design, yet they often determine
practical cycle life.

Current generative AI primarily optimizes the initial crystal structure,
whereas practical batteries require optimization of the evolving material.
Future inverse-design frameworks should therefore incorporate
electro-chemo-mechanical durability directly into the generative process by
considering insertion-induced strain, anisotropic deformation, stress
localization, fracture susceptibility, stress-dependent diffusion, interfacial
mechanics, and temperature-dependent structural evolution. The objective is no
longer simply to generate stable materials, but to design materials whose
chemistry, crystallography, transport, and mechanics remain favorably coupled
throughout repeated electrochemical cycling.

\subsection{Tier 6: Electrode and cell compatibility}
\label{sec:tier6}
\TierGoal{Generate cell-compatible electrodes.}

The preceding tiers focus primarily on the intrinsic properties of the active
material. However, even a porous oxide with excellent thermodynamic
stability, transport characteristics, electrochemical functionality, and
mechanical durability cannot deliver practical battery performance unless it
functions effectively within a complete electrode and cell architecture.
Battery operation emerges from the coupled interactions among the active
material, electrolyte, conductive additives, binder, current collector,
separator, and counter electrode. The objective of the sixth tier is therefore
to optimize electrode and cell compatibility, recognizing that battery
performance is fundamentally a systems-level property rather than an
intrinsic material property.

Among these interactions, the electrolyte plays a central role. Beyond
transporting ions between electrodes, it governs desolvation, interfacial
charge transfer, electrochemical stability, and formation of the solid
electrolyte interphase (SEI, \cref{fig:sei}). Consequently, the performance of
a porous framework depends not only on its crystallographic channels but also
on the chemistry and mechanics of the electrode--electrolyte interface. Future
AI-driven discovery should therefore treat the active material and electrolyte
as a coupled design problem rather than optimizing them independently.

Practical electrodes introduce additional engineering constraints. Particle
morphology, surface chemistry, adhesion, packing density, conductive
additives, binders, and mass loading collectively determine electronic
connectivity, electrolyte infiltration, volumetric energy density, and
long-term structural integrity. These factors demonstrate that crystal
structure alone cannot predict electrode performance. The architecture of the
composite electrode must also be considered.

At the full-cell level, further phenomena, including dendrite formation,
thermal stability \citep{chatterjee2020nanoporous}, safety, and current distribution, emerge from
the coupled behavior of both electrodes, the electrolyte, interfaces
\citep{sharma2021variation,sharma2021understanding,basu2018utilizing}, and operating conditions. Future inverse-design
frameworks should therefore evolve beyond optimizing individual materials
toward designing integrated electrode--electrolyte systems capable of reliable
operation under realistic conditions.

This tier represents an important conceptual transition. Earlier generations
of generative AI primarily optimized crystal structures as isolated materials.
Future AI should instead optimize functional battery systems, where materials,
interfaces, electrolytes \citep{hasan2025computational}, electrode architectures, and operating
conditions are designed simultaneously. Only through this systems-level
perspective can computationally generated porous oxides be translated into
deployable energy-storage technologies.

\subsection{Tier 7: Manufacturability and sustainability}
\label{sec:tier7}
\TierGoal{Generate deployable technologies.}

The preceding tiers establish whether a porous oxide can function as an
effective battery material. The final tier asks a different question: can it
be deployed at scale \citep{attia2025challenges}? A material that exhibits outstanding
computational performance but cannot be synthesized economically,
manufactured reproducibly, or deployed sustainably is unlikely to contribute
meaningfully to future energy-storage technologies. The objective of the
seventh tier is therefore to incorporate manufacturability and sustainability
directly into the inverse-design process, ensuring that AI discovers not only
scientifically promising materials but also deployable technologies.

Practical deployment depends on engineering, economic, and supply-chain
considerations that extend beyond intrinsic electrochemical performance.
Elemental abundance, resource availability, toxicity, and precursor
accessibility influence scalability and environmental impact, while synthesis
conditions, including processing temperature, reaction atmosphere, process
complexity, and energy consumption, govern manufacturing cost and
reproducibility. Likewise, synthetic yield, phase purity, and process
robustness determine whether computationally attractive materials can be
translated into reliable laboratory synthesis and ultimately industrial
production. Future generative frameworks should therefore evaluate these
constraints alongside conventional electrochemical descriptors rather than
treating them as post-discovery considerations.

Sustainability introduces an equally important dimension. The environmental
footprint of a battery material depends not only on its composition but also
on its complete life cycle, including raw-material extraction, synthesis,
battery fabrication, operation, recycling, and end-of-life management. Scarce
elements, energy-intensive processing, or low synthetic yield may
substantially increase cost and carbon footprint even when electrochemical
performance appears exceptional. Future AI-driven materials discovery should
therefore optimize scientific performance together with economic viability and
environmental responsibility.

This tier completes the progression from crystal discovery to technology
deployment. Earlier tiers focused on chemistry, physics, transport,
electrochemistry, durability, and battery-system integration. The final step
recognizes that successful battery materials must also satisfy manufacturing,
economic, and societal constraints. The optimal porous oxide is therefore not
necessarily the material with the highest theoretical capacity or lowest
migration barrier, but the one that best balances electrochemical
performance, durability, manufacturability, cost, and sustainability.

Current generative AI largely treats these considerations as external
constraints applied after materials discovery. Future autonomous discovery
systems should instead incorporate them directly into the generative process,
enabling candidate materials to be optimized simultaneously from scientific,
engineering, economic, and environmental perspectives. This completes the
transition from generic crystal generation to physics-informed,
application-aware, and deployment-driven inverse design, where AI no longer
discovers only promising materials but identifies those most likely to become
practical energy-storage technologies.

\section{The Missing Data Problem and Autonomous Knowledge Generation}
\label{sec:missingdata}

\subsection{Beyond crystal databases: the Missing Data Problem}
\label{sec:beyond-db}

\begin{figure}[tbp]
  \centering
  \includegraphics[width=0.64\linewidth]{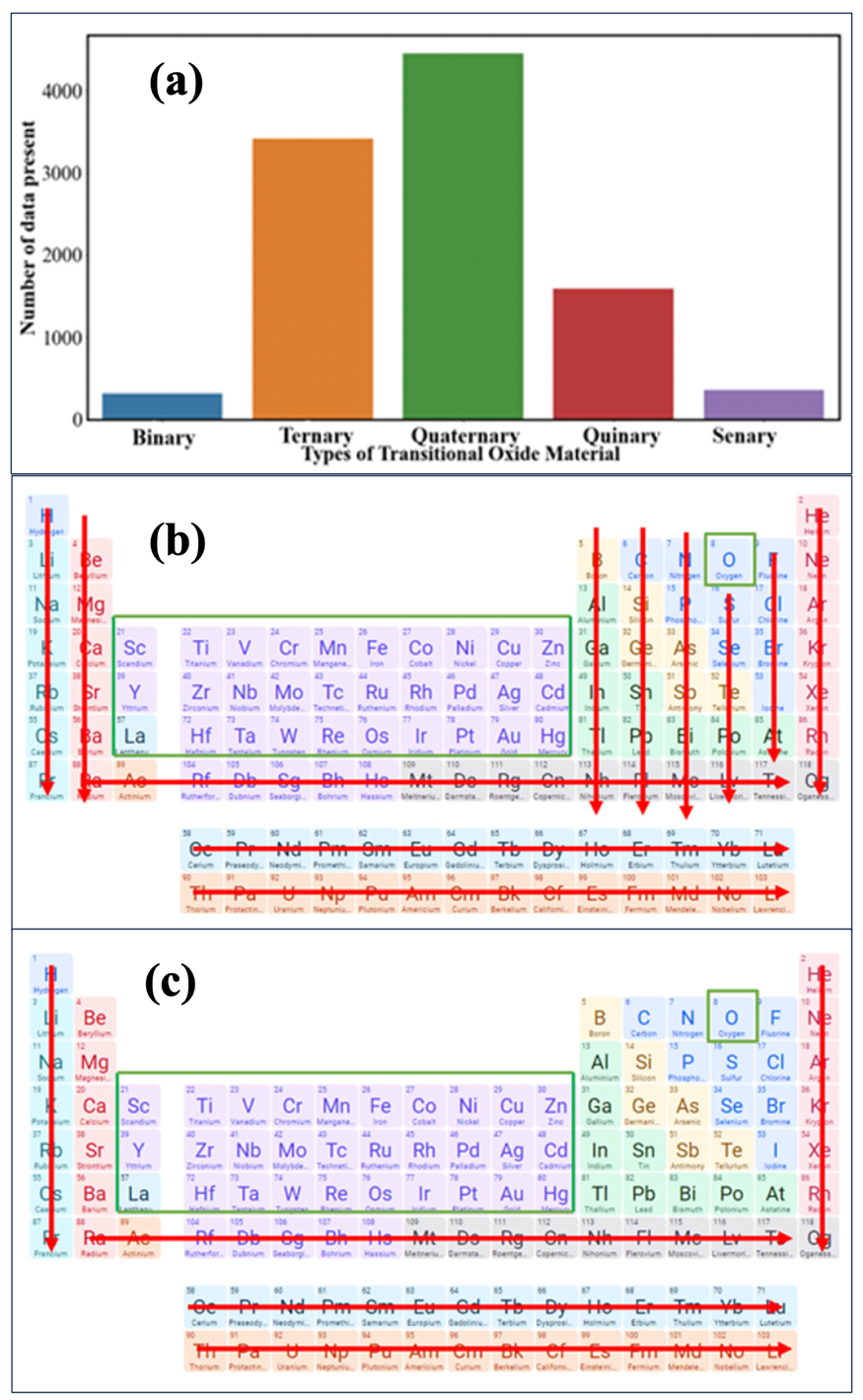}
  \caption{Data distribution of transition-metal oxide (TMO)-based materials.
  (a) Number of data entries available for each class of TMO composition,
  from binary to senary. (b) Overview of binary TMOs. (c) Overview of
  ternary-based TMOs. In (b) and (c), green represents the considered
  materials and red depicts materials ruled out from our list. Areas without
  color in (c) represent the third element in ternary-based TMOs. Binary TMOs (b)
  combine a transition metal with oxygen. In the ternary-based TMOs (c),
  multicomponent TMO materials
  consist of elements from any element group except group-1 alkali metals,
  aligning with our focus on multivalent ions. For example, quaternary-based
  TMOs follow the format W$_s$A$_r$B$_m$O$_n$, where W represents any metal
  from the periodic table excluding monovalent metals, A represents any metal
  from the periodic table, B represents the transition metal, and $s$, $r$,
  $m$, and $n$ denote the stoichiometric ratios. The possible TMO structures are therefore
  effectively unlimited, whereas existing databases contain very limited
  data. Reprinted with permission from ref.~\citep{datta2024unlocking}. Copyright
  2024 Royal Society of Chemistry.}
  \label{fig:tmodata}
\end{figure}

The remarkable progress of generative artificial intelligence in materials
science has been driven largely by comprehensive computational materials
databases. Resources such as the Materials Project \citep{jain2013materials}, the Open
Quantum Materials Database (OQMD) \citep{kirklin2015oqmd}, AFLOW \citep{curtarolo2012aflow}, ICSD
\citep{hellenbrandt2004icsd}, and related initiatives have transformed computational materials
discovery by providing millions of crystal structures together with
first-principles predictions of formation energy, electronic structure,
elastic properties, phase stability, and other fundamental descriptors. These
databases established the foundation for high-throughput screening, machine
learning, and, more recently, generative models capable of proposing
chemically plausible crystal structures. The success of first-generation
generative AI is therefore inseparable from these databases
(\cref{fig:tmodata}a). By learning statistical relationships among
composition, crystal symmetry, topology, and computed properties, diffusion
models, crystal variational autoencoders, and large language models have
demonstrated an unprecedented ability to generate novel crystal structures
beyond existing crystallographic repositories. However, these databases were
developed primarily to support computational materials science rather than
application-aware materials design.

For porous oxide electrodes, practical battery performance depends on far more
than crystal structure and equilibrium thermodynamics. It requires knowledge
of ion-specific transport, electrochemical behavior, electro-chemo-mechanical
durability, synthesis conditions, electrode architecture, electrolyte
compatibility, manufacturability, and sustainability. These descriptors are
either absent or only sparsely represented in existing databases
(\cref{fig:tmodata}a), even though the space of possible compositions is
effectively infinite (\cref{fig:tmodata}b,c). Instead, they remain dispersed
throughout thousands of scientific publications using inconsistent
terminology, reporting conventions, experimental protocols, and units
\citep{birkle2020web,burnham2006scopus,ginsparg2011arxiv,haddaway2015role} (\cref{fig:framework}). This challenge extends
beyond conventional data scarcity. The required scientific knowledge already
exists, but much of it remains inaccessible to artificial intelligence because
it is embedded within unstructured scientific literature rather than organized
into machine-readable knowledge.

\begin{KeyIdea}[The Missing Data Problem]
We refer to this limitation as the \emph{missing data problem} \citep{datta2026autonomous}.
The problem is therefore not the absence of knowledge, but the absence of a
structured representation that AI can readily learn from.
\end{KeyIdea}

The implications become particularly evident in the context of the seven-tier
inverse-design framework proposed in the previous section. Designing battery
materials that simultaneously satisfy chemical validity, transport
functionality, electrochemical performance, electro-chemo-mechanical
durability, manufacturability, and sustainability requires knowledge that
extends far beyond crystallographic databases. Without access to these
multidimensional relationships, even the most sophisticated generative models
remain limited to optimizing only a subset of the properties governing
practical battery performance.

The next bottleneck in autonomous materials discovery is therefore unlikely
to be model architecture alone, but the availability of sufficiently rich
scientific knowledge to support application-aware inverse design \citep{datta2026autonomous}.
Addressing the missing data problem requires more than expanding existing
databases: it demands a new knowledge infrastructure capable of integrating
crystallography, electrochemistry, mechanics, synthesis, manufacturing, and
experimental evidence into a unified, machine-readable representation. Such an
infrastructure is a prerequisite for the next generation of autonomous
materials discovery (\cref{fig:framework}).

\subsection{From heterogeneous literature to machine-readable knowledge}
\label{sec:literature}

Scientific knowledge describing porous oxide materials already exists in
enormous quantities, but it remains dispersed across heterogeneous sources
rather than organized into machine-readable databases. A typical battery
materials study simultaneously reports crystal structures, synthesis
procedures, electrochemical measurements, transport properties, degradation
mechanisms, and mechanistic interpretation through narrative text, tables,
figures, equations, supplementary information, computational datasets, and
experimental metadata. These complementary representations are designed for
human communication rather than automated reasoning, while differences in
terminology, units, experimental protocols, and reporting conventions make
direct integration across independent studies inherently difficult.

Addressing the Missing Data Problem therefore requires far more than
digitizing publications or expanding existing databases. The central challenge
is to transform heterogeneous scientific information into structured,
machine-readable knowledge while preserving its physical meaning,
experimental context, and scientific provenance. We refer to this process as
\emph{autonomous knowledge generation}, in which AI continuously acquires,
extracts, harmonizes, and integrates scientific evidence into a form suitable
for autonomous reasoning and inverse design.

\begin{figure}[tbp]
  \centering
  \includegraphics[width=0.70\linewidth]{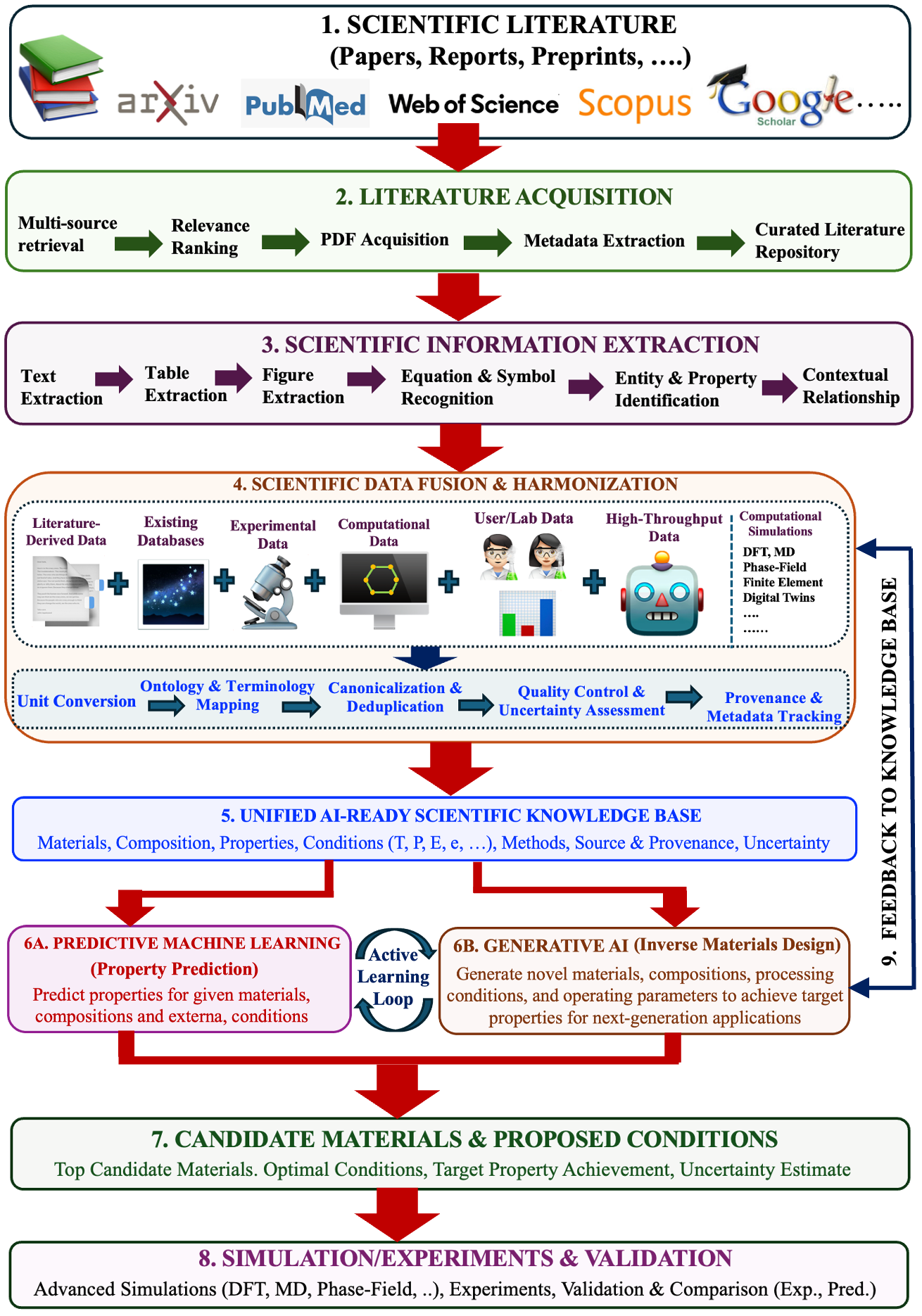}
  \caption{An autonomous scientific knowledge generation framework for
  AI-driven scientific discovery \citep{datta2026autonomous}.}
  \label{fig:framework}
\end{figure}

This process consists of tightly connected stages \citep{datta2026autonomous}
(\cref{fig:framework}). Ontology-guided literature acquisition identifies
scientifically relevant publications based on semantic concepts rather than
isolated keywords. Multimodal information extraction recovers complementary
information from text, tables, figures, equations, supplementary information,
and metadata. Semantic harmonization reconciles differences in terminology,
units, reporting conventions, and experimental context so that observations
from independent studies become directly comparable. Finally, knowledge fusion
integrates these harmonized observations into coherent scientific
representations while preserving provenance, uncertainty, and confidence.

The resulting framework differs fundamentally from conventional literature
mining. Rather than producing isolated summaries or collections of extracted
values, autonomous knowledge generation constructs an interconnected
scientific knowledge network in which crystallography, transport,
electrochemistry, mechanics, synthesis, and manufacturing become semantically
linked. Such a representation enables AI not only to retrieve information but
also to reason across scientific disciplines, identify previously hidden
relationships, and continuously incorporate new scientific evidence as it
becomes available.

Autonomous knowledge generation therefore provides the missing bridge between
the scientific literature and the application-aware inverse-design framework
proposed in \cref{sec:tiers}. As future generative AI systems increasingly
optimize multidimensional objectives spanning chemistry, transport,
electrochemistry, electro-chemo-mechanics, synthesis, and sustainability,
their performance will depend as much on the quality of the underlying
scientific knowledge as on advances in model architecture. Rather than
functioning as another literature-mining pipeline, autonomous knowledge
generation establishes the continuously evolving knowledge foundation required
for the next generation of autonomous materials discovery.

\subsection{A Porous Oxide Energy Materials Ontology}
\label{sec:ontology}

The autonomous knowledge-generation framework described above requires a
common semantic representation capable of integrating crystallographic,
electrochemical, mechanical, synthesis, and manufacturing knowledge from
heterogeneous scientific sources. We propose a \emph{Porous Oxide Energy
Materials Ontology}, which defines the entities, relationships, and
descriptors needed to represent porous oxide electrodes within a unified
semantic framework. Rather than serving as a controlled vocabulary, the
ontology enables autonomous AI systems to organize, interpret, and reason
across the multidimensional relationships governing battery performance
throughout the complete materials development cycle.

The ontology is organized into six interconnected semantic layers describing
material identity, pore architecture, electrochemical context and
performance, electro-chemo-mechanical behavior, synthesis and
manufacturability, and supporting evidence (\cref{tab:ontology}). Together,
these layers provide a comprehensive representation of porous oxide materials
that extends beyond crystal structure alone.

\paragraph{Material identity} establishes the structural foundation of the
ontology through descriptors including composition, crystal structure,
polymorph, space group, oxidation states, substitutions, defect chemistry,
and crystallographic disorder.

\paragraph{Pore architecture} describes the defining structural features of
porous oxides through channel geometry, accessible and limiting pore
diameters, bottleneck size, connectivity, tortuosity, transport
dimensionality, and ion-specific accessibility.

\paragraph{Electrochemical context and performance} links materials behavior
to operating conditions by recording the charge carrier, electrolyte, voltage
window, temperature, current density, C-rate, electrode loading, cell
configuration, and corresponding performance metrics such as voltage,
gravimetric and volumetric capacity, Coulombic efficiency, energy density,
power density, rate capability, and cycle life.

\paragraph{Electro-chemo-mechanical behavior} captures the coupled processes
governing long-term durability, including insertion-induced strain, stress
evolution, elastic properties, volume expansion, fracture resistance, phase
transformation, amorphization, and other degradation mechanisms.

\paragraph{Synthesis and manufacturability} connect computational prediction
with experimental realization through descriptors such as precursor
chemistry, synthesis route, processing conditions, particle morphology,
crystallinity, yield, and phase purity.

\paragraph{Evidence.} Finally, the evidence layer preserves provenance,
methodology, uncertainty, confidence, and publication metadata for every
extracted observation, allowing AI systems to evaluate evidence quality,
reconcile conflicting reports, and propagate uncertainty throughout
downstream reasoning.

\begin{table}[tbp]
  \caption{The six semantic layers of the proposed Porous Oxide Energy
  Materials Ontology (\cref{sec:ontology}), the descriptors each records, and
  the role each plays in the seven-tier inverse-design hierarchy of
  \cref{sec:tiers}.}
  \label{tab:ontology}
  \centering
  \small
  \begin{tabular}{L{0.21\linewidth} L{0.50\linewidth} L{0.21\linewidth}}
    \toprule
    \DDHeadRow
    \DDHead{Semantic layer} & \DDHead{Representative descriptors} & \DDHead{Role in the hierarchy} \\
    \midrule
    Material identity &
      composition, crystal structure, polymorph, space group, oxidation
      states, substitutions, defect chemistry, crystallographic disorder &
      establishes chemical validity \\
    Pore architecture &
      channel geometry, accessible and limiting pore diameters, bottleneck
      size, connectivity, tortuosity, transport dimensionality, ion-specific
      accessibility &
      defines transport functionality \\
    Electrochemical context and performance &
      charge carrier, electrolyte, voltage window, temperature, current
      density, C-rate, loading, cell configuration; voltage, capacity,
      Coulombic efficiency, energy and power density, rate capability,
      cycle life &
      captures battery operation \\
    Electro-chemo-mechanical behavior &
      insertion-induced strain, stress evolution, elastic properties, volume
      expansion, fracture resistance, phase transformation, amorphization,
      degradation mechanisms &
      characterizes durability \\
    Synthesis and manufacturability &
      precursor chemistry, synthesis route, processing conditions, particle
      morphology, crystallinity, yield, phase purity &
      links computational discovery with experimental realization \\
    Evidence &
      provenance, methodology, uncertainty, confidence, publication metadata &
      ensures scientific traceability \\
    \bottomrule
  \end{tabular}
\end{table}

More importantly, the ontology mirrors the seven-tier inverse-design hierarchy
introduced in \cref{sec:tiers}. Material identity establishes chemical
validity; pore architecture defines transport functionality; electrochemical
context captures battery operation; electro-chemo-mechanical descriptors
characterize durability; synthesis links computational discovery with
experimental realization; and the evidence layer ensures scientific
traceability. The ontology therefore functions not simply as a data schema,
but as the semantic backbone of application-aware generative AI, enabling
autonomous discovery systems to reason consistently across multiple
scientific disciplines within a unified scientific framework.

\subsection{Toward a continuously evolving Porous Oxide Energy Materials Knowledge Base}
\label{sec:kb}

The ontology described in the previous subsection provides the semantic
foundation for organizing scientific knowledge. The next step is to transform
this semantic framework into a continuously evolving \emph{Porous Oxide
Energy Materials Knowledge Base}. Unlike conventional materials databases,
which are periodically updated through manual curation, this knowledge base is
envisioned as a living scientific resource that continuously acquires,
integrates, and refines knowledge as new computational and experimental
evidence becomes available. Its purpose is not merely to store information,
but to establish the knowledge infrastructure required for autonomous
materials discovery.

The knowledge base should integrate complementary evidence from multiple
scientific sources. Peer-reviewed literature provides synthesis strategies,
mechanistic understanding, electrochemical characterization, and experimental
observations. First-principles calculations \citep{datta2024unlocking} contribute
atomic-scale information describing crystal structures, thermodynamic
stability, electronic properties, diffusion pathways, and defect energetics.
Molecular dynamics \citep{kim2014atomistic,fan2013mechanical} and phase-field simulations
\citep{steinbach2009phase,wang2020application} extend these descriptions to finite-temperature transport,
mesoscale evolution, stress development, and degradation, while experiments
provide essential validation of computational predictions. Looking further
ahead, autonomous laboratories equipped with robotic synthesis and closed-loop
experimentation could continuously generate standardized data that both
validate AI-generated hypotheses and expand the knowledge base.

The strength of this framework lies not in maintaining separate repositories
of literature, simulations, and experiments, but in knowledge fusion.
Ontology-guided integration semantically connects complementary observations
from different sources, allowing computational predictions to be compared
directly with experiments, conflicting evidence to be reconciled, uncertainty
to be propagated, and independent observations to be combined into
progressively more complete descriptions of materials behavior. The result is
an interconnected scientific knowledge network that supports both human
interpretation and autonomous reasoning.

Equally important, the proposed knowledge base is designed for continuous
learning. As new publications appear, computational methods improve, and
experimental techniques reveal previously inaccessible phenomena, the
knowledge infrastructure continuously incorporates new evidence, updates
confidence estimates, and refines semantic relationships among materials,
properties, synthesis conditions, and performance metrics. The knowledge base
therefore evolves together with the scientific understanding it represents.

Such a knowledge infrastructure is essential for the next generation of
application-aware generative AI. Future inverse-design systems must
simultaneously optimize crystallographic stability, transport functionality,
electrochemical performance, electro-chemo-mechanical durability,
manufacturability, and sustainability, an interconnected knowledge landscape
that no existing materials database can adequately represent. A continuously
evolving knowledge base therefore becomes not merely a computational resource,
but the scientific foundation for autonomous inverse design.

\begin{KeyTakeaway}
The proposed Porous Oxide Energy Materials Knowledge Base should therefore be
viewed not simply as another materials database, but as a living scientific
knowledge infrastructure. By continuously integrating literature,
simulations, experiments, and autonomous laboratories within a common
semantic framework, it provides the knowledge backbone required for
physics-informed, application-aware generative AI. In this emerging paradigm,
scientific knowledge is no longer the endpoint of discovery; it becomes the
continuously evolving foundation upon which autonomous materials discovery is
built.
\end{KeyTakeaway}

\section{Synthesis-Aware and Closed-Loop Autonomous Discovery}
\label{sec:closedloop}

\subsection{From crystal generation to synthesis-aware design}
\label{sec:synthesis-aware}

Generative AI has dramatically accelerated exploration of crystallographic
design space, yet crystal generation represents only the first step toward
practical materials discovery. A newly generated porous oxide acquires
scientific value only when it can be synthesized, characterized, and
experimentally validated. The next evolution of inverse design must therefore
extend beyond crystal generation toward synthesis-aware materials discovery
\citep{kwade2018current}, where computational prediction and experimental realization
become tightly integrated.

One promising direction is the incorporation of scientific knowledge directly
into the design process through retrieval-augmented reasoning. Rather than
relying solely on statistical relationships learned during training, future
AI systems should retrieve synthesis strategies, precursor chemistry, reaction
conditions, characterization protocols, and mechanistic insights from the
scientific literature to guide experimental planning. In this way, the
literature becomes an active component of inverse design rather than merely a
source of background information. However, synthesis planning is substantially
more complex than crystal generation. Successful synthesis depends not only on
precursor selection but also on reaction thermodynamics, competing phases,
kinetics, processing atmosphere, heating schedules, equipment, and other
practical factors that remain poorly represented in current AI models.

\subsection{Toward closed-loop autonomous discovery}
\label{sec:closed-loop}

The long-term vision extends beyond synthesis-aware materials discovery toward
a closed-loop autonomous discovery framework, in which materials generation,
computational validation, synthesis planning, experimentation, and knowledge
generation continuously inform one another rather than operating as
independent stages (\cref{fig:framework}). Within this framework,
physics-informed generative AI first proposes candidate porous oxides
constrained by chemistry, crystallography, transport, and electrochemical
requirements. Predictive machine-learning models then perform
multi-objective evaluation of thermodynamic stability, electrochemical
performance, electro-chemo-mechanical durability, manufacturability, and
uncertainty, enabling Pareto optimization \citep{jin2008pareto,petchrompo2022review} across competing
objectives. The most promising candidates undergo progressively
higher-fidelity validation through first-principles calculations, multiscale
simulations, and ontology-guided knowledge retrieval to identify plausible
synthesis strategies. Active learning subsequently prioritizes the most
informative experiments, while robotic synthesis and high-throughput
characterization progressively automate experimental execution.

\begin{KeyIdea}[Experimentation is not the endpoint]
The defining feature of this framework is that experimentation is not the
endpoint of discovery. Every successful synthesis, failed experiment,
unexpected phase, and discrepancy between prediction and observation becomes
new scientific evidence that is incorporated into the continuously evolving
knowledge infrastructure introduced in \cref{sec:missingdata}. Computational
prediction, experimentation, and knowledge generation therefore become tightly
coupled components of a self-improving scientific discovery system.
\end{KeyIdea}

\section{Challenges and Research Roadmap}
\label{sec:roadmap}

The rapid progress of generative artificial intelligence has demonstrated that
autonomous materials discovery is no longer a speculative concept but an
emerging scientific reality. However, realizing the vision proposed throughout
this perspective requires advances extending far beyond model architecture.
Future progress will depend equally on improvements in chemical reliability,
application-aware optimization, autonomous experimentation, scientific
governance, and knowledge infrastructure. Rather than representing independent
challenges, these priorities define a research roadmap spanning the near,
medium, and long term.

\subsection{Near-term priorities: making generative AI chemically controllable}
\label{sec:nearterm}

The immediate priority is to improve the chemical reliability of generative
models. Although current crystal generators successfully explore
crystallographic design space, they frequently violate fundamental chemical
constraints, as illustrated by the oxygen deficiency observed in
first-generation porous oxide generation \citep{datta2025generative}. Future models should
therefore incorporate chemistry directly into the generative process through
explicit compositional constraints, charge neutrality, oxidation-state
consistency, coordination chemistry, and oxide-aware decoding rather than
relying on post-generation filtering.

Progress also requires richer structural representations. Instead of treating
crystals solely as periodic atomic arrangements, future models should
explicitly encode pore geometry, channel connectivity, bottleneck dimensions,
transport dimensionality, and ion-specific accessibility as intrinsic design
variables. Equally important are reliable uncertainty quantification,
standardized benchmarks, reproducible workflows, and openly accessible
datasets that enable rigorous comparison among competing generative
approaches while improving confidence in AI-assisted materials discovery.

\subsection{Medium-term priorities: making generative AI application-aware}
\label{sec:mediumterm}

Once chemically reliable generation becomes established, the next objective is
to transform crystal generation into application-aware inverse design. Future
AI systems must simultaneously optimize multiple competing objectives,
including electrochemical performance, electro-chemo-mechanical durability,
manufacturability, cost, safety, and sustainability, rather than focusing
primarily on thermodynamic stability. Multi-objective Pareto optimization will
therefore become essential for navigating scientifically meaningful trade-offs
among these competing design criteria.

Application-aware design also requires moving beyond equilibrium crystal
structures toward realistic battery operation. Future inverse-design
frameworks should explicitly incorporate ion-specific pore engineering,
finite-temperature stability, electrolyte compatibility, interphase evolution,
electro-chemo-mechanical degradation, and electrode-level performance. Equally
important, synthesis feasibility should become an intrinsic design objective
rather than a post-discovery validation step, enabling AI to prioritize
materials that are both scientifically promising and practically realizable.

\subsection{Long-term vision: autonomous laboratories and self-improving discovery}
\label{sec:longterm}

The long-term vision extends beyond computational design toward fully
integrated autonomous discovery ecosystems. Advances in robotics,
high-throughput characterization, active learning, and autonomous
experimentation are creating the foundation for laboratories capable of
proposing, synthesizing, characterizing, and refining new materials with
minimal human intervention.

Within such systems, every experiment becomes new scientific knowledge.
Successful syntheses validate predictive models, while failed syntheses,
impurity formation, and unexpected phase evolution reveal the limits of
current understanding and expose previously unknown scientific phenomena.
Future knowledge infrastructures should therefore preserve both positive and
negative experimental outcomes, enabling continuously evolving AI systems to
refine their scientific understanding as new evidence becomes available.

In this paradigm, materials discovery evolves into a self-improving scientific
ecosystem in which computation, experimentation, and knowledge generation
continuously reinforce one another. Rather than operating as separate
activities, they become tightly integrated components of an autonomous
discovery process capable of accelerating materials innovation far beyond what
either computation or experimentation can achieve independently.

\section{Conclusions}
\label{sec:conclusions}

Generative artificial intelligence is transforming materials discovery by
shifting computational research from forward prediction toward inverse design.
Recent advances in diffusion models, variational autoencoders, large language
models, and related generative architectures have demonstrated that AI can
efficiently explore vast crystallographic design spaces and generate
chemically plausible materials beyond existing databases. These developments
represent a major milestone in computational materials science. However, for
porous oxide electrodes, crystal generation alone represents only the first
step toward practical materials discovery.

Throughout this perspective, we have argued that the next generation of
AI-enabled materials discovery must move beyond generating stable crystal
structures toward application-aware inverse design. The performance of porous
oxide electrodes emerges from coupled interactions among crystal chemistry,
pore architecture, ion transport, electrochemistry, electro-chemo-mechanics,
synthesis, manufacturing, and battery-system integration. Future AI systems
must therefore learn these multidimensional relationships and incorporate them
directly into the generative process rather than treating them as independent
post-generation screening criteria.

To support this transition, we proposed a seven-tier physics-informed
inverse-design framework that progressively integrates chemical validity,
thermodynamic viability, transport functionality, electrochemical performance,
electro-chemo-mechanical durability, cell compatibility, and manufacturability
into a unified design hierarchy. Rather than representing independent
evaluation criteria, these tiers define a roadmap for evolving generative AI
from crystal generation toward the design of experimentally realizable and
technologically deployable battery materials.

We further argued that realizing this vision requires solving the Missing Data
Problem. Existing crystallographic databases enabled the first generation of
generative AI by providing crystal structures and first-principles properties,
but application-aware inverse design requires substantially richer scientific
knowledge describing synthesis, transport, electrochemistry, mechanics,
manufacturing, and experimental evidence. Addressing this challenge demands
autonomous knowledge generation, ontology-guided semantic integration, and
continuously evolving knowledge infrastructures capable of transforming
heterogeneous scientific literature into machine-readable scientific
knowledge.

Building upon this concept, we introduced the Porous Oxide Energy Materials
Knowledge Base as a living scientific knowledge infrastructure rather than a
conventional materials database. By integrating literature, multiscale
simulations, experiments, and autonomous laboratories within a unified
semantic framework, such a knowledge base provides the foundation for
Synthesis-Aware, Closed-Loop Autonomous Discovery, where computational design,
experimental validation, and knowledge generation become tightly coupled
components of a continuously learning scientific system.

Although this perspective focuses on porous oxide materials for
next-generation energy storage, the concepts developed here extend well beyond
this application. The seven-tier inverse-design framework, ontology-driven
knowledge infrastructure, and synthesis-aware closed-loop discovery paradigm
together provide a general blueprint for AI-enabled materials discovery across
a broad range of functional materials, including solid electrolytes,
electrocatalysts, quantum materials, structural alloys, polymers, and
biomaterials.

Ultimately, we envision a future in which artificial intelligence is no longer
viewed simply as a tool for generating candidate materials, but as an active
scientific partner participating throughout the entire discovery cycle, from
knowledge acquisition and inverse design to computational modeling,
experimentation, and continual learning. The next frontier of AI-enabled
materials science therefore lies not in developing larger generative models,
but in building integrated scientific ecosystems in which knowledge
generation, computation, experimentation, and reasoning operate
synergistically. Realizing this vision will require sustained collaboration
among materials scientists, chemists, physicists, experimentalists, computer
scientists, and AI researchers. Such interdisciplinary efforts have the
potential to transform materials discovery from a largely sequential process
into a continuously learning, knowledge-driven enterprise capable of
accelerating the development of next-generation sustainable materials.

\begin{Declarations}
\Statement{Lead contact}{%
  Further information and requests for resources and materials should be
  directed to and will be fulfilled by the author, Dibakar Datta
  (\href{mailto:ddlab@njit.edu}{ddlab@njit.edu}).}
\Statement{Materials availability}{No materials were synthesized in this study.}
\DataAvailability{%
  Data can be obtained by request from the author. Any additional information
  required to reanalyze the data reported in this paper is available from the
  author upon request.}
\CodeAvailability{%
  Complete code for the case study in \cref{sec:casestudy} is available at
  \url{https://github.com/joy1303125/Generative-AI-for-battery-material}.
  Test code showing how to implement autonomous scientific discovery (data
  fusion), as in \cref{fig:framework}, is available (with the author's
  permission) at
  \url{https://drive.google.com/drive/folders/1DDPBGw_eTWViJedSPIf20a462bwJYz_z?usp=drive_link}.}
\CompetingInterests{The author has no conflict of interest to declare.}
\Funding{%
  This work is supported by the National Science Foundation (NSF) CAREER Award
  (award number 2237990).}
\Acknowledgments{%
  The author acknowledges the Advanced Cyberinfrastructure Coordination
  Ecosystem: Services \& Support (ACCESS) for the computational facilities
  (award numbers DMR180013 and MAT250009) used for the case study presented in
  \cref{sec:casestudy}.}
\end{Declarations}

\setlength{\bibsep}{1.8pt plus 0.6pt}
\bibliographystyle{unsrtnat}
\bibliography{references}

\end{document}